\documentclass[11pt,a4paper]{article}

\usepackage{amsmath,amsthm,amssymb}
\usepackage{graphics,graphicx}
\usepackage{epsfig}
\usepackage{multicol}
\usepackage{color}
\makeatletter
\@addtoreset{equation}{section}
\makeatother

\newcommand{\M}{\mathcal M}

\begin{document}
\vspace{0.5cm}
\begin{center}
\Large{\bf Generalised DBI-Quintessence}
\end{center}
\begin{center}
\large{\bf Burin Gumjudpai} ${}^{a, b, }$\footnote{buring@nu.ac.th, B.Gumjudpai@damtp.cam.ac.uk} \large{\bf and John Ward}
${}^{c,}$\footnote{jwa@uvic.ca}
\end{center}
\begin{center}
\emph{ ${}^a$ Fundamental Physics \& Cosmology Research Unit, The Tah Poe Academia Institute (TPTP)\\ Department of Physics, Naresuan
University, Phitsanulok 65000, Siam, Thailand}
\\
\vspace{0.5cm} \emph{${}^b$ Centre for Theoretical Cosmology, DAMTP, University of Cambridge \\ CMS, Wilberforce Road, Cambridge, CB3 0WA, U.K.}
\\
\vspace{0.5cm}
\emph{${}^c$ Department of Physics and Astronomy \\University of Victoria, Victoria, BC, V8P 1A1, Canada}\\
\end{center}
\begin{abstract}
We investigate the phase space of a quintessence theory governed by a generalised version of the DBI action, using a combination of numeric and
analytic methods. The additional degrees of freedom lead to a vastly richer phase space structure, where the field covers the full equation of
state parameter space; $-1 \le \omega \le 1$. We find many non-trivial solution curves to the equations of motion which indicate that DBI
quintessence is an interesting candidate for a viable k-essence model.
\end{abstract}
\section{Introduction}
The dark energy problem continues to be a sticking point for theoretical physicists. The simplest solution to this problem is to
postulate the existence of a vacuum energy or cosmological constant which agrees with all the current observational bounds
\cite{Spergel:2006hy, Komatsu:2008hk, Riess:1998cb, Tegmark:2003ud}.
However we are then left with a
secondary problem, namely explaining why the vacuum energy is tuned to such a small value without some obvious symmetry to protect it.
For many years we have hoped that UV complete theories of gravity would shed light on this issue, which is in effect an extremely
embarrassing IR problem from this perspective. However despite much effort, neither string theory nor loop quantum gravity has shed any compelling
light on this issue
 - although there have been many interesting proposals.

An alternative approach is to assume that the cosmological constant is exactly zero, since supersymmetry can then be invoked as the regulating
symmetry in this case. However one then has to account for the fact that low-energy supersymmetry must be broken and an alternative explanation
for the current expansion must be sought. One way to deal with the latter problem is to assume that the dark energy phase is driven by a
dynamical field, implying that the equation of state is an explicit function of time \cite{Copeland:2006wr, Peebles:1987ek}. Currently this cannot be ruled out by our best
observations and therefore remains a possible solution to the dark energy problem. However one cannot just consider ad hoc scalar fields
coupling to gravity, since the low energy theory will still be sensitive to high energy physics. In particular we must ensure that any
additional scalars are neutral under all the standard model symmetries, and that they do not introduce additional fifth forces. Therefore one
must search for viable models of dynamical dark energy within UV sensitive theories.

Phenomenological models of our universe have proven difficult to construct within string theory, due to technical difficulties arising from
moduli stabilization, whereby we assume that the extra dimensions of the theory are compactified on manifolds with $SU(3)\times SU(3)$ structure (in the type IIB case)
\cite{Grana:2005jc}, and orientifolded to preserve the minimal amount of supersymmetry in four-dimensions. As a result, embedding realistic
cosmology into string theory has proven difficult. One area which has been well explored in recent years, is inflation driven by the open string
sector through dynamical $Dp$-branes. This is the so-called DBI inflation \cite{Alishahiha:2004eh, Kecskemeti:2006cg} - which lies in a special class of
K-inflation models. It was originally thought that such models yielded large levels of non-Gaussian perturbations which could be used as a
falsifiable signature of string theory \cite{Maldacena:2002vr}. However subsequent work has shown that this is may not be the case, and
that the simplest DBI models are essentially indistinguishable from standard field theoretic slow roll models \cite{Lidsey:2007gq, Alabidi:2008ej, Huston:2008ku} \footnote{Note however, that the models proposed in \cite{Avgoustidis:2006zp, Huston:2008ku} evade such
problems.}. The problem is that the WMAP $5$ year data set \cite{Komatsu:2008hk} imposes very tight constraints on the allowed tuning of the free
parameters in the theory. We are then left with the choice of either having large non-Gaussianities but with vanishing tensors, or assume that
the tensor spectrum will be visible - in which case there is no non-Gaussian signature. The models are only falsifiable once these conditions
are relaxed. One can get around these conditions by considering more complicated models such as multiple fields \cite{Easson:2007dh}, multiple branes
\cite{Cai:2008if, Cai:2009hw, Thomas:2007sj}, wrapped branes \cite{Becker:2007ui} or monodromies \cite{Silverstein:2008sg} - but even here there are still problems with
fine tuning, backreaction and the apparent breakdown of perturbation theory in the inflationary regime \cite{Leblond:2008gg}. 

In models of dynamical dark energy, on the other hand \cite{Copeland:2006wr, Peebles:1987ek, ArmendarizPicon:2000dh},
the WMAP constraints can be relaxed and therefore DBI models may
still have some use as an explanation for a dynamical equation of state. Moreover this fits in nicely with several intuitive ideas from string
theory. Namely that inflation can still occur, albeit only through the closed string sector - where one (or more) of the geometric pseudo-moduli
are actually responsible for the initial inflationary epoch (see \cite{Conlon:2005jm} for the phenomenologically most viable proposals). After
inflation the universe lives on branes that wrap various cycles within the compact space and are extended along the large Minkowski directions.
In this sense we see that a GUT or Electro-Weak (EW) phase transition can manifest through a geometric fashion - namely the Higgsing of branes
in the compact bulk space. This suggests that dark energy may well be a dynamical process, and moreover in the light of these open string
constructions, retains a sense of being geometric in nature.

With this in mind, various authors have begun to explore the phase space of DBI-driven dark energy \cite{Martin:2008xw, Guo:2008sz}. The initial works have dealt with the
dynamics of a solitary $D3$-brane moving through a particular warped compactification of type IIB. In this note we wish to generalise this
further to a more phenomenological class of models that include multiple and partially wrapped branes. We believe that this may be a more
generic situation to consider, since typically one should expect branes of varying degrees to be wrapped on non-trivial cycles of the compact
space. Our work is a first step into considerations of a more general set-up for quintessence in IIB (open)-string theory, and we hope will be a
valuable starting point for further endeavour.
\section{Dynamics of the effective theory}
To begin let us assume that the the universe at such late times can be adequately described by a flat FRW metric and the matter sector consists of a dynamical scalar field
and a perfect fluid, which are both separately conserved. The usual cosmological equations of motion are therefore independent of any particular model
and can be written as
\begin{eqnarray}
H^2 &=& \frac{(\rho + \rho_{\phi})}{3 M_p^2}\,, \\
\dot{\rho_i}&=&- 3 H (P_i + \rho_i)\,, \nonumber
\end{eqnarray}
where $i$ runs over the contributing components. The equation of state is given by $\omega_i = P_i/\rho_i$, however if $\omega$ of the fluid
component is assumed to be constant then we can integrate the appropriate conservation equation exactly to obtain
\begin{equation}
\rho \propto a^{-3(1+\omega)}\,,
\end{equation}
where the scale factor varies as a function of time such that $a(t) \sim t^{2/(3[1+\omega])}$.

The model dependence arises in the parameterisation of the scalar field sector.
In our case we are assuming that the dark energy is driven by open string modes, which at low energies are described by fluctuations of a $Dp$-brane
whose dynamics are governed by the Dirac-Born-Infeld action (DBI) - which is a generalisation of non-linear electrodynamics \cite{Alishahiha:2004eh, Kecskemeti:2006cg}.
Typically one assumes that the standard model is localised on an intersecting brane stack, in one of the many warped throats that are attached to the internal
space. For consistency reasons in the simplest cases, these are taken to be either $D3$ or $D7$-branes. In this note we will consider a bottom up approach
therefore we shall not worry too much about the geometric deformations of the compact space, nor any constraints imposed by Orientifold $Op$-planes - aside
from those that ensure that all tadpoles are consistently canceled so that we can trust the low energy supergravity theory.

The action we consider is a generalised form of the DBI one coupled to Einstein-Hilbert gravity, which can be embedded into this background and takes
the following generalised form \footnote{We refer the more interested readers to \cite{Thomas:2007sj} for more details on the precise structure and origin of this action. The important thing to note is that $\phi$ is a matrix valued field. For recent work in a related direction see \cite{Ashoorioon:2009wa}.}
\begin{equation}
S = -\int d^4 x a^3(t) \left( T(\phi)W(\phi)\sqrt{1-\frac{\dot{\phi}^2}{T(\phi)}}-T(\phi) + \tilde V(\phi) \right) + S_M\,,
\end{equation}
where $T(\phi)$ is the warped tension of the brane and $S_M$ is the action for matter localised in the Standard Model (SM) sector. Thus
our assumption here is that our dynamical open string sector is coupled only gravitationally to the SM sector and so we do not have to worry about
additional forces or particle production.
There are two potential terms for the scalar field which are denoted by $W(\phi)$ and $\tilde V(\phi)$. The
first of these terms can arise in different places within the theory. Firstly if the brane is actually a non-BPS one \cite{Sen:2002in}, then the scalar field mode is actually
tachyonic and the potential is therefore of the usual runaway form. If there are $N$ multiple coincident branes, then the world-volume field theory is a $U(N)$
non-Abelian gauge theory and the potential term is simply a reflection of the additional degrees of freedom \cite{Myers:1999ps}.
Through the dielectric effect, one can also
see that this configuration is related to a $D5$-brane wrapping a two-cycle within the compact space and carrying a non-zero magnetic flux along this cycle. Both
of these configurations lead to an additional potential multiplying the usual DBI kinetic term.

The origin of the $\tilde V(\phi)$ term is less explicit - but is a sum of terms.
One expects open or closed string interactions to generate a scalar potential $V(\phi)$, however the precise
form of such an interaction depends upon many factors such as the number of additional branes and geometric moduli,
the number of non-trivial cycles in the compact space, and the choice of embedding for branes on these cycles.
Typically one can only compute this in special cases in the full string theory. There are also additional terms coming from coupling of the brane to any background RR form
fields. The action above is assumed to be that of $D3$-brane(s) filling the space-time directions, which naturally couple to the field $C^{(4)}$ through
the Chern-Simons part of the action. However for wrapped $D5$-branes there is also the possibility of a coupling $C^{(4)} \wedge F$, where $F$ is the magnetic field
through the two-cycle. For example in the warped deformed conifold one can see that $d C^{(6)} = \star d C^{(2)}$ and therefore there is an additional term in
the DBI action
\begin{equation}
S \sim \int d^4 x a^3(t) g_s^{-1} M \alpha' T(\phi)\,,
\end{equation}
up to a normalisation factor of order one. Terms such as this have been added to the interaction potential to define the \emph{full} scalar
potential $\tilde V(\phi)$. Recent extensions to standard DBI inflation have included the contribution from higher dimensional bulk forms, with the remarkable
result that they cancel one another up to third order in the action and therefore do not affect the leading order perturbations \cite{Langlois:2009ej}. Extending this work to higher
order is therefore extremely interesting.

The corresponding equations for the energy density and pressure of the DBI can then be written succinctly as
\begin{equation}
P_{\phi} = \frac{T(\phi)}{\gamma} \left[\gamma - W(\phi)\right]  - \tilde V(\phi)\,, \hspace{0.5cm} \rho_{\phi}= T(\phi)\left[W(\phi)\gamma-1
\right] + \tilde V(\phi)\,,
\end{equation}
where $\gamma = [1-\dot{\phi}^2/T(\phi)]^{-1/2}$ is the usual generalisation of the relativistic factor. The subscript $_\phi$ denotes the
scalar field component here. We can also immediately define the equation of state parameter for the quintessence field to be
\begin{equation}
\omega_{\phi} = \frac{T(\phi)\left[\gamma-W(\phi)\right]-\tilde V(\phi) \gamma}{T(\phi)\gamma\left[W(\phi)\gamma-1\right]+\tilde
V(\phi)\gamma}\,,
\end{equation}
from which one clearly sees that it is dynamically sensitive and can take a wide range of values.
For instance we only recover $\omega_{\phi}\sim -1$ in the limit that the field is
non-relativistic and the entire solution is dominated by the $\tilde V(\phi)$ terms - which will clearly require large amounts of fine tuning to
accomplish. There are clearly several regions of parameter space that are of interest. First let us assume that the potential term is zero,
either because it is suppressed or there is an unlikely cancellation between the contributing terms. The more general case with non-zero $\tilde
V$ leads to a wide variety of complex behaviour. We can therefore identify several limits of interest - focusing on the behaviour of $W$:
\begin{itemize}
\item $W(\phi)=1$ - which reduces the action back to the usual DBI case which has $\omega_{\phi} = 1/ \gamma$ as discussed in \cite{Martin:2008xw}.
\item $W(\phi) = \alpha \gamma$ - which leads to constant $\gamma$ if $\alpha$ is constant,
since the two are related via $\gamma^2 \omega_{\phi} \alpha = 1-\alpha+\omega_{\phi}$. Moreover this again means that $\dot{\phi} \propto
t^{-(1+\omega_{\phi})/(1+\omega)}$ as in the case where $W=1$.
\item $W(\phi) \to 0$ - as could occur in the case of a tachyonic theory, which mimics a dark energy dominates phase with $\omega_{\phi}=-1$. However
one must be careful if this is to be representative of non-BPS $D$-brane actions, since the coupling to the form field is non canonical in this
instance. In fact the coupling term will typically be of the form $d\phi \wedge C$. This means that there is no solitary $T(\phi)$ term in the
action and therefore the equation of state in this instance will vary like $-1/\gamma^2$.
\item $W(\phi) \gg \gamma$ - which can occur in the multi-brane/wrapped brane case and yields $\omega_{\phi} \sim - 1/ \gamma^2$.
\end{itemize}
Note that in all cases the equation of state parameter remains bounded between $-1 \le \omega_{\phi} \le 1$.

One can combine the expressions for the energy-momentum tensor components, and together with
the continuity equation we obtain the following equation of motion - assuming that the scalar field follows a monotonic path
\begin{equation}
\ddot{\phi}+\frac{3H\dot{\phi}}{\gamma^2}+\frac{3 T_{\phi}}{2\gamma^2} + \frac{1}{W\gamma^3}(\tilde
V_{\phi}-T_{\phi})-\frac{T_{\phi}}{2}+\frac{T W_{\phi}}{W\gamma^2}=0\,,
\end{equation}
which is a generalisation of the Klein-Gordon equation for the DBI Lagrangian. The subscript $_{\phi}$ of $T, W$ and $\tilde V$ denotes
derivative with respect to the field value. The other dynamical equation of motion for the Hubble parameter can be written as
\begin{equation}
\dot{H} = -\frac{1}{2 M_p^2} \left[\rho(1+\omega) + \gamma W(\phi) \dot{\phi}^2 \right]\,,
\end{equation}
where we have defined the pressure of the barotropic fluid to be $P = \omega \rho$ and that it is non-interacting. We leave the interesting case
of interacting pressure for future endeavour.

Let us consider, as an example solution, the case where there is a scaling solution with $W=1$, which has been reviewed elsewhere \cite{Guo:2008sz}.
We will find it convenient to define the
quantity
\begin{equation}
X = \frac{1+\omega_{\phi}}{1+\omega}\,,
\end{equation}
in which case we see that $\dot{\phi} \sim t^{-X}$. This allows us to reconstruct the tension of the brane as follows
\begin{eqnarray}
T(\phi) &=& \M^4 e^{-\lambda \phi}\,, \hspace{2.3cm} X=1\\
&=& \M^{4+\alpha} \phi^{-\alpha}\,, \hspace{2cm} X \ne 1 \nonumber
\end{eqnarray}
where $\M$ is a dimensionful mass scale, $\lambda$ is a constant and $\alpha = 2X/(1-X)$. Using the fact that $\omega_{\phi} = 1/\gamma$ we can then see that
for $X \ne 1$ the solution is physically valid only when $\omega > 2/\alpha$ since we define $\gamma$ to be the positive root.
Let us now consider the phase-space dynamics of the theory in more detail following along the lines of \cite{Copeland:2006wr}. It is initially
convenient to define the following new variables
\begin{eqnarray}
x &=& \sqrt{\frac{T(\phi)W(\phi)\gamma}{3}}\frac{1}{H M_p}, \hspace{0.5cm} \mu_1 = \frac{\sqrt{T}M_p \tilde V_{\phi}}{\tilde{V}^{3/2}}\,, \nonumber \\
y &=& \sqrt{W(\phi)\gamma} \frac{\dot{\phi}}{H M_p}\,, \hspace{1.5cm} \mu_2 = - \frac{\sqrt{T}M_p T_{\phi}}{\tilde V^{3/2}}\,, \nonumber \\
z &=& \sqrt{\frac{\tilde{V}}{3}}\frac{1}{H M_p}\,, \hspace{2.3cm} \mu_3 = \frac{W_{\phi}M_p}{W^{3/2}\gamma^{5/2}}\,,
\end{eqnarray}
in terms of which we can see that $\gamma = [1-y^2/(3x^2)]^{-1/2}$ and the fluid density parameter can be written as
\begin{equation}
\Omega = 1 - \Omega_{\phi} = 1- \left(z^2 + x^2 \left[1-\frac{1}{W(\phi)\gamma} \right] \right)\,,
\end{equation}
whilst the equation of state in dimensionless variables will become
\begin{equation}
\omega_{\phi} = \frac{1}{\gamma} \left(\frac{x^2[\gamma-W(\phi)]-z^2 W(\phi) \gamma^2}{x^2 [W(\phi) \gamma-1] +z^2 W(\phi) \gamma} \right).
\end{equation}
As is customary we will now switch to dimensionless derivatives, denoted by a prime, replacing time derivatives by derivatives with respect to
the e-folding number, $\mathcal{N}$. Therefore we can easily determine
\begin{equation}
\frac{H'}{H} = -\frac{y^2}{2} - \frac{3(1+\omega)}{2} \left(1-z^2-x^2 \left[1-\frac{1}{W(\phi)\gamma} \right] \right)\,. \label{Hdash}
\end{equation}
A useful quantity to calculate is the variation of the kinetic function, which we can write in the following manner using the equation of motion
\begin{equation}
\frac{\dot{\gamma}}{\gamma} = -\frac{3H \dot{\phi}^2}{T}-\frac{W_{\phi} \dot{\phi}}{W}-\frac{T_{\phi} \dot{\phi}}{T}-\frac{\dot{\phi}}{\gamma W
T}(\tilde V_{\phi}-T_{\phi})\,.
\end{equation}
 We can then determine the dynamical equations for the dimensionless fields as derivatives with respect to $\mathcal{N}$
\begin{eqnarray}
x' &=& -\frac{1}{2}(\mu_1 + \mu_2) \frac{y z^3}{x^2}-\frac{y^2}{2x}- x\frac{H'}{H}\,, \nonumber \\
y' &=& - 3y \left(1-\frac{y^2}{6x^2} \right)\left(1+\frac{z^3}{xy}[\mu_1+\mu_2] \right) + \frac{3 \mu_2 z^3 W}{\gamma x}-3x^2\mu_3 - y \frac{ H'}{H}\,, \nonumber \\
z'&=& \frac{z^2y \mu_1}{2x} - z \frac{ H'}{H}\,, \label{zdash}
\end{eqnarray}
and the remaining parametric solutions are
\begin{eqnarray}
\mu_1' &=& \frac{\mu_1^2 y z}{x}\left(-\frac{3}{2}+\frac{\tilde V_{\phi \phi}\tilde V}{\tilde V_{\phi}^2}+\frac{T_{\phi}\tilde V}{\tilde V \tilde V_{\phi}} \right)\,, \nonumber \\
\mu_2' &=& \frac{\mu_1\mu_2yz}{x}\left(-\frac{3}{2}+\frac{T_{\phi}\tilde V}{2T \tilde V_{\phi}}+ \frac{T_{\phi \phi}\tilde V}{T_{\phi} \tilde V_{\phi}} \right)\,, \nonumber \\
\mu_3' &=& y \mu_3^2 \gamma^{3/2}\left(1+ \frac{W_{\phi \phi}W}{W_{\phi}^2}+ \frac{5 T_{\phi}W}{2TW_{\phi}}+\frac{5}{2T\gamma W_{\phi}}[\tilde
V_{\phi}-T_{\phi}] \right) + \frac{5 \mu_3 y^2}{2x^2}\,.   \label{mudash}
\end{eqnarray}
Note that if the $\mu_i$ are constants, then the previous three equations form an autonomous set and should uniquely specify the dynamics of the
quintessence field. We will consider this case as the simplest (canonical) example. If we wish to appeal to string theoretic constructions then
we restrict the parameter space of solutions. It is more interesting to consider the above equations in the context of a phenomenological model
and see what kind of functions yield the correct behaviour. Explicit constructions of string backgrounds are typically difficult and there are
only a few well known examples that are ritually invoked, however if we take string theory seriously then there are undoubtedly other
non-trivial backgrounds that are cosmologically interesting but not yet constructed. Since an analytic analysis of this generalised system is
highly complicated, it is convenient to use a combination of analytic and numerical methods to understand the dynamics of the system. For a
numeric analysis it is necessary to re-write the fluid equation in terms of more useful variables. It turns out that the simplest variables to
use are the following
\begin{eqnarray}
\phi' &=& \Phi \label{phi}\,, \\
\Phi' &=& -\frac{3 \Phi}{\gamma^2}+ \frac{3 M_p z^3}{x} \left(\frac{\sqrt{W \gamma} \mu_2}{2}\left\lbrack\frac{3}{\gamma^2}-1\right\rbrack-
\frac{(\mu_1+\mu_2)}{\sqrt{W}\gamma^{5/2}} \right) - \frac{3 M_p x^2 \mu_3}{\sqrt{W\gamma}}-\Phi \frac{H'}{H} \,, \label{Phi}
\end{eqnarray}
which are easily derivable from the terms written above. The equations (\ref{Hdash}), (\ref{zdash}), (\ref{mudash}), (\ref{phi}) and (\ref{Phi})
together with barotropic fluid equation: $\rho' = - 3 \rho(\mathcal{N}) (1+w)$, hence form a closed ten-dimensional autonomous system if $T, W$
or $\tilde V$ are given as explicit functions of $\phi$ or as constants.
\subsection{Case I}
Let us take the canonical string theoretic example arising when the local geometry can be approximated by an $AdS$ space. This geometry
typically arises in the near horizon limit of coincident $D3$-branes (or flux). In this case we see that (at leading order)
\begin{equation}
T(\phi) = \frac{\phi^4}{\lambda^4}\,, \hspace{1cm} \tilde{V}(\phi)=\frac{m^2\phi^2}{2}\,, \hspace{1cm} W(\phi)=W\,,
\end{equation}
where we have also included an effective $\phi^2$ potential for the system. This means that $\mu_3=0$ and we also have a constant $\mu_1$ which allows us to
write the remaining $\mu$ terms as
\begin{equation}
\mu_1 = \frac{2\sqrt{2} M_p}{m \lambda^2}\,, \hspace{1cm} \mu_2= -\frac{2x^2 \mu_1}{W\gamma z^2}\,,
\end{equation}
and therefore the dynamical equations reduce to
\begin{eqnarray}
x'&=& -\frac{\mu_1yz^3}{2x^2}\left(1-\frac{2x^2}{W\gamma z^2} \right)-\frac{y^2}{2x}-\frac{x H'}{H}\,, \nonumber \\
y' &=& - 3y \left(1-\frac{y^2}{6x^2} \right)\left(1+\frac{z^3 \mu_1}{xy}\left \lbrack 1-\frac{2x^2}{W \gamma z^2}\right \rbrack \right) - \frac{6 \mu_1 zx}{\gamma^2}- \frac{yH'}{H}\,, \nonumber \\
z'&=& \frac{z^2y\mu_1}{2x}-\frac{zH'}{H}\,.
\end{eqnarray}
The simplest way to proceed with the analysis is to consider the final equation above, since this splits the solution space neatly into two
components. Thus we search for solutions where either $z=0$ or $z = (2x/y \mu_1) H'/H$ as initial conditions.

The first sub-set of solutions admits $(0,0,0)$ as a (trivial) fixed point, which is a fluid dominated solution since $\Omega=1$ in this
instance. Let us remark here that this fixed point solution will occur for \emph{all} the cases we consider, however since this implies a
vanishing of the action, causality implies that this fixed point must be unstable - i.e. phase space trajectories will flow away from it. By
making this field a phantom scalar, one can evade this causal bound and the point can become a stable fixed point. This behaviour arises in many
places in the literature, so we will not discuss it further here.

There is also a critical point at $(1,\sqrt{3},0)$ which is a kinetic dominated solution. This solution actually exists as solutions to the
quadratic expression $y^2=3 x^2$ which corresponds to the limit $\gamma \to \infty$. In terms of the density parameter, a quick calculation
shows that along the general curve (parameterised by $y_0$ and $x_0$) we find $\Omega = 1-x_0^2$. Thus at the trivial fixed point we see $\Omega
\to 1$, however for $x_0 \to 1$ we see that $\Omega \to 0$ corresponding to non-relativistic matter, i.e. dust. In this instance we also find
$a(t) \sim t^{2/3}$ as expected from the cosmological evolution equations. Again due to the special algebraic properties of the DBI action, we
anticipate that this solution will also be found for the other cases of interest.

The second sub-set of solutions are more interesting, as initially one can solve the system by slicing the phase space at $y=0$
\footnote{Note that one cannot do this for $x=0$ since the action becomes singular and ill-defined.}.
One can use the condition on $H'$ to
fix $z$ through $z^2=1-x^2(W-1)/W$. Combining this with the equations of motion gives us the following fixed point (taking positive signs of all roots for simplicity)
\begin{equation}
x=\sqrt{\frac{W}{1-W}}\,, \hspace{0.5cm} y=0\,, \hspace{0.5cm} z= 1\,,
\end{equation}
which is valid for all $W < 1$ in order for these points to be real and at finite distance in phase space. If we then compute the density of the
fluid we find $\Omega = 0$, since $\Omega_{\phi}=1$, which corresponds to a purely dust-like solution. Note that this class of solutions does
not exist for the simple $D3$-brane analysis as in \cite{Guo:2008sz}, since it arises from additional degrees of freedom which are neglected in
these models. The remaining solutions in this sub-set are difficult to find analytically.

More generally we can see that the above solution corresponds is a special case of the more general Case I behavior, which we
paramaterise by
\begin{equation}
T(\phi) = \frac{\phi^{\alpha}}{\lambda^{\alpha}}\,, \hspace{1cm} \tilde V(\phi) = \frac{m^\beta \phi^\beta}{\beta}\,, \hspace{1cm} W(\phi)=W\,,
\end{equation}
where we can then explicitly write
\begin{equation}
\mu_1 = A \left(\frac{x}{z \gamma^{1/2}} \right)^{(\alpha-\beta-2)/(\alpha-\beta)}\,, \hspace{0.5cm} \mu_2 = - \frac{\alpha}{\beta}
\frac{\mu_1}{W\gamma} \frac{x^2}{z^2}\,, \hspace{0.5cm} \mu_3 =0\,,
\end{equation}
where $A$ is a (real, positive) constant provided that $\beta > 0$.
\begin{equation}
A = \frac{M_p \beta^{3/2}}{\lambda^{\alpha/2} m^{\beta/2}}\left(\frac{\lambda^{\alpha} m^{\beta}}{\beta W}
\right)^{(\alpha-\beta-2)/2(\alpha-\beta)}\,,
\end{equation}
but which simplifies in the limit $\alpha=\beta+2$. As before, the solution space splits into two disconnected sub-sets, therefore in the first
instance where we take slices through $z=0$, we find the following bound
\begin{equation}
\frac{2}{(\alpha-\beta)}>0\,,
\end{equation}
which implies that $\alpha>\beta$ and so the brane tension should dominate the dynamics (in the large field regime).
Let us therefore assume that $\alpha, \beta$ are chosen such that
this condition is satisfied - then we find the solution branch is governed again by the relation $y^2 = 3x^2$ as expected - which contains the solution $(0,0,0)$ as
a special case. Moreover this is valid for all
values of $\alpha, \beta$ satisfying the above constraint.
The secondary solution branch occurs when we find solutions to
\begin{equation}
\frac{zy \mu_1}{2x}= \frac{H'}{H}\,,
\end{equation}
which is generally very complicated. A simple set of solutions do arise when we consider slices at $y=0$, since the fixed points are localised along the curve
\begin{equation}
x = \pm \sqrt{\frac{\beta W}{(\alpha-\beta)(1-W)}}, \hspace{0.5cm} y=0, \hspace{0.5cm} z = \pm \sqrt{\frac{\alpha}{(\alpha-\beta)}}\,,
\end{equation}
which corresponds to a dust-like solution $\Omega = 0$ $\forall \alpha, \beta$.
The reality constraint here demands that $\alpha > \beta$ which in turn
fixes $W <1$. However there are also additional solutions where $\beta <0$ and positive $\alpha$ - provided that $W > 1$. Explicit realisations of this scenario
within a string theory context can arise through potentials arising from brane/anti-brane interactions and is therefore a non-trivial and interesting
solution.

\begin{figure}
\begin{center}
\includegraphics[width=3.0in]{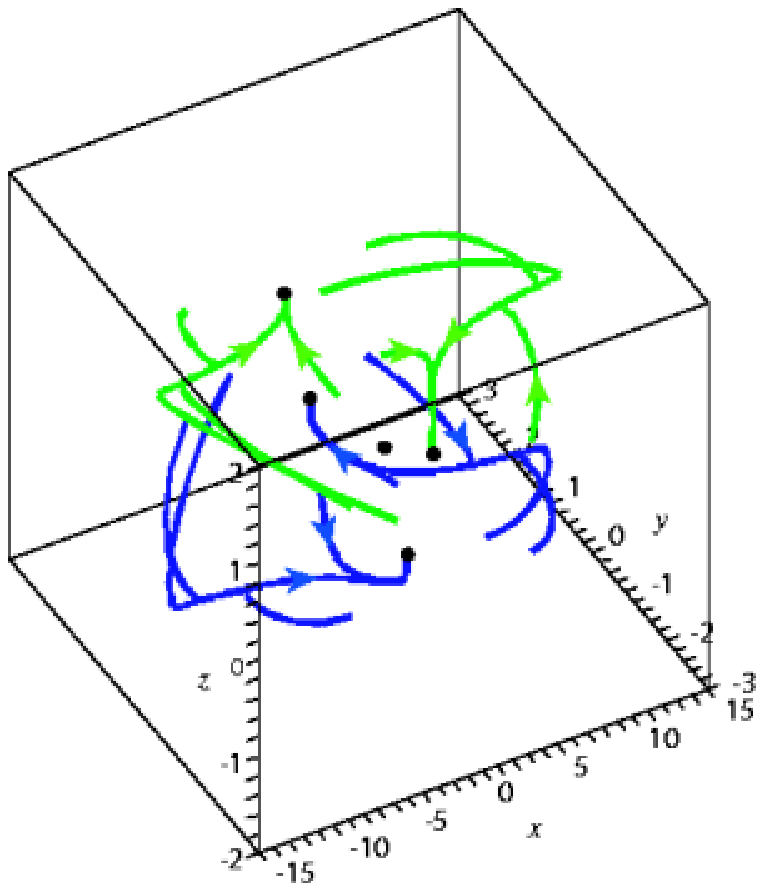}
\caption{(Case I) 3-D $xyz$ phase space trajectories for $T(\phi) = \phi^4/\lambda^4, \tilde V(\phi) = m^2 \phi^2 /2$ and $W(\phi) =W$. We have
set here, $M_p =1, W=1, m=1, \lambda =1$ and $w = 0$ (dust case). } \label{figcase1_xyz}
\end{center}
\vspace{0.5cm}
\begin{center}
\includegraphics[width=2.7in]{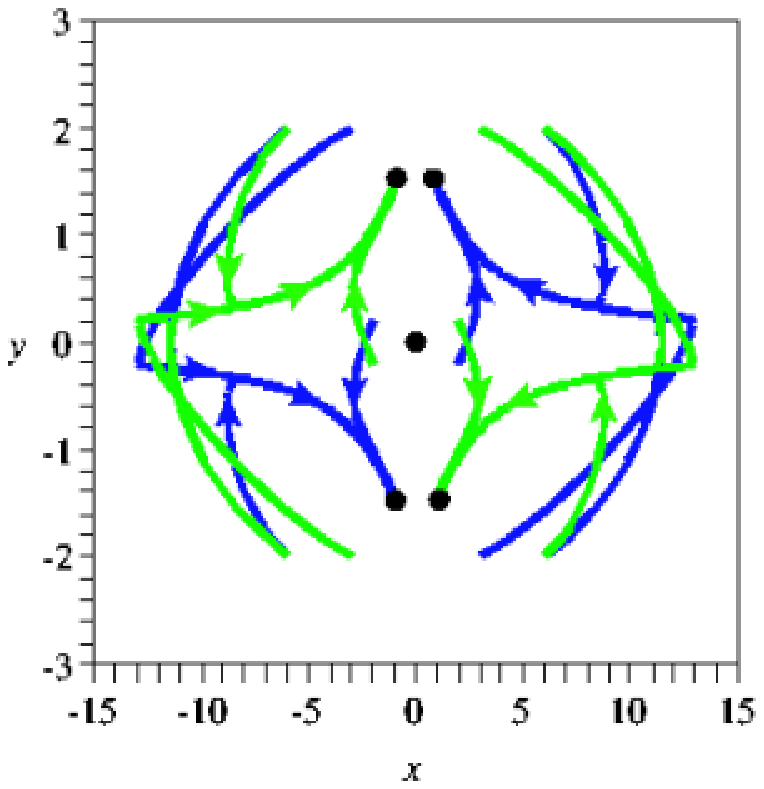}
\caption{(Case I) Phase space trajectories in $xy$ plane. Four attractors $(\pm 1, \pm \sqrt{3}, 0)$ and one unstable node (0, 0, 0) can be seen
here. $z$ is bounded within $(-1,1)$ range.} \label{figcase1_xy}
\end{center}
\end{figure}

\begin{figure}
\begin{center}
\includegraphics[width=2.7in]{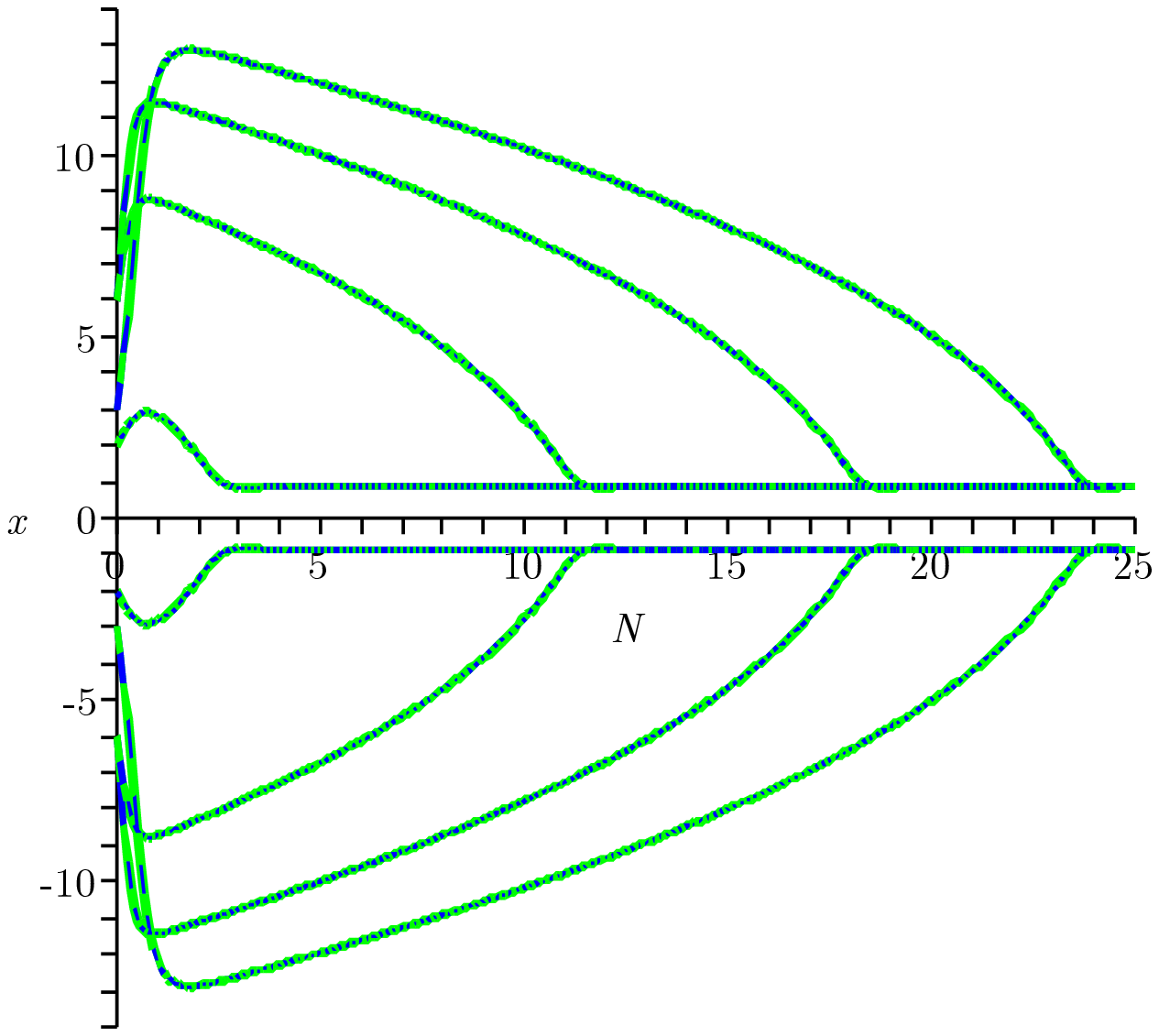}
\end{center}
\begin{center}
\includegraphics[width=2.7in]{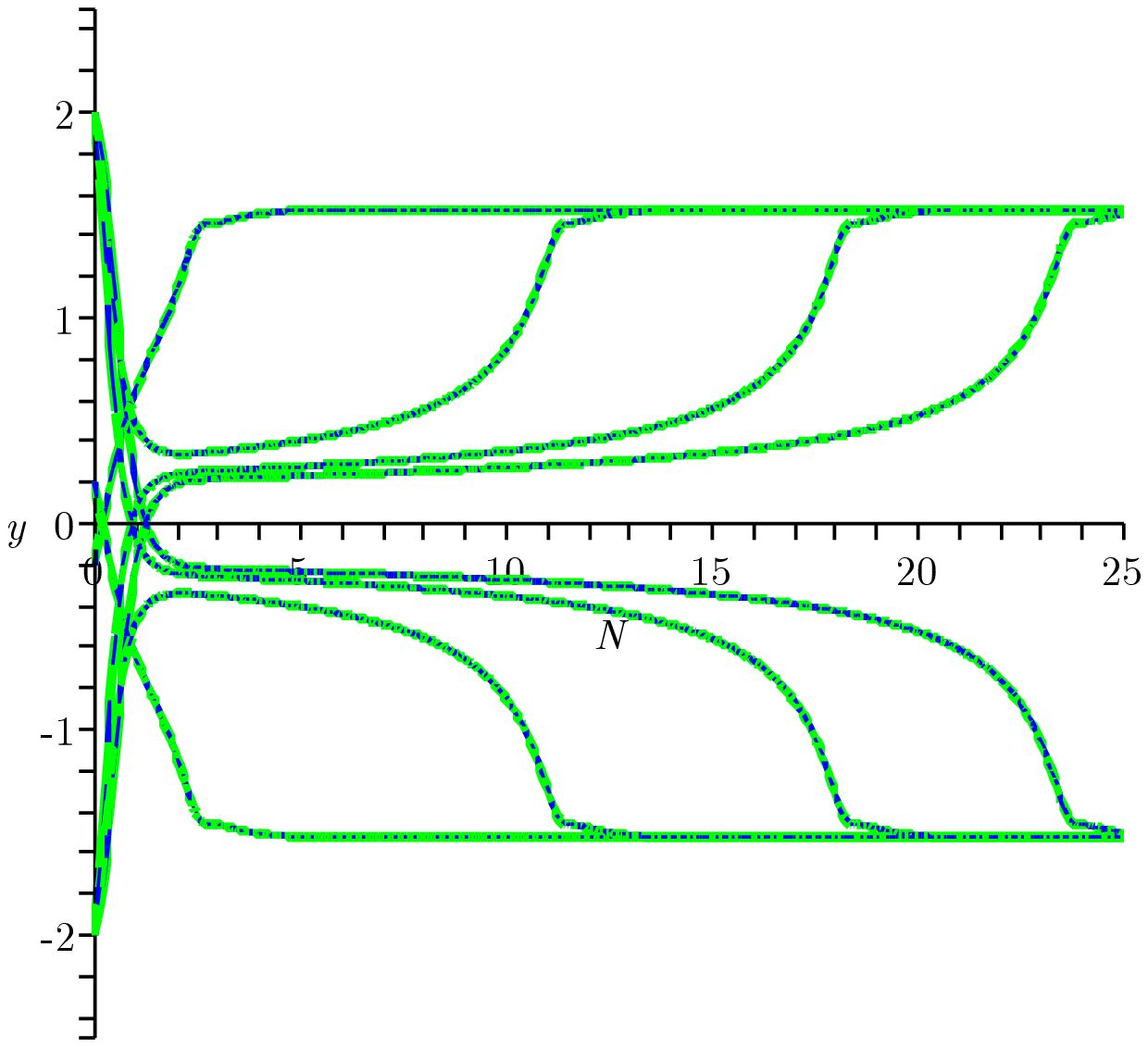}
\end{center}
\begin{center}
\includegraphics[width=2.7in]{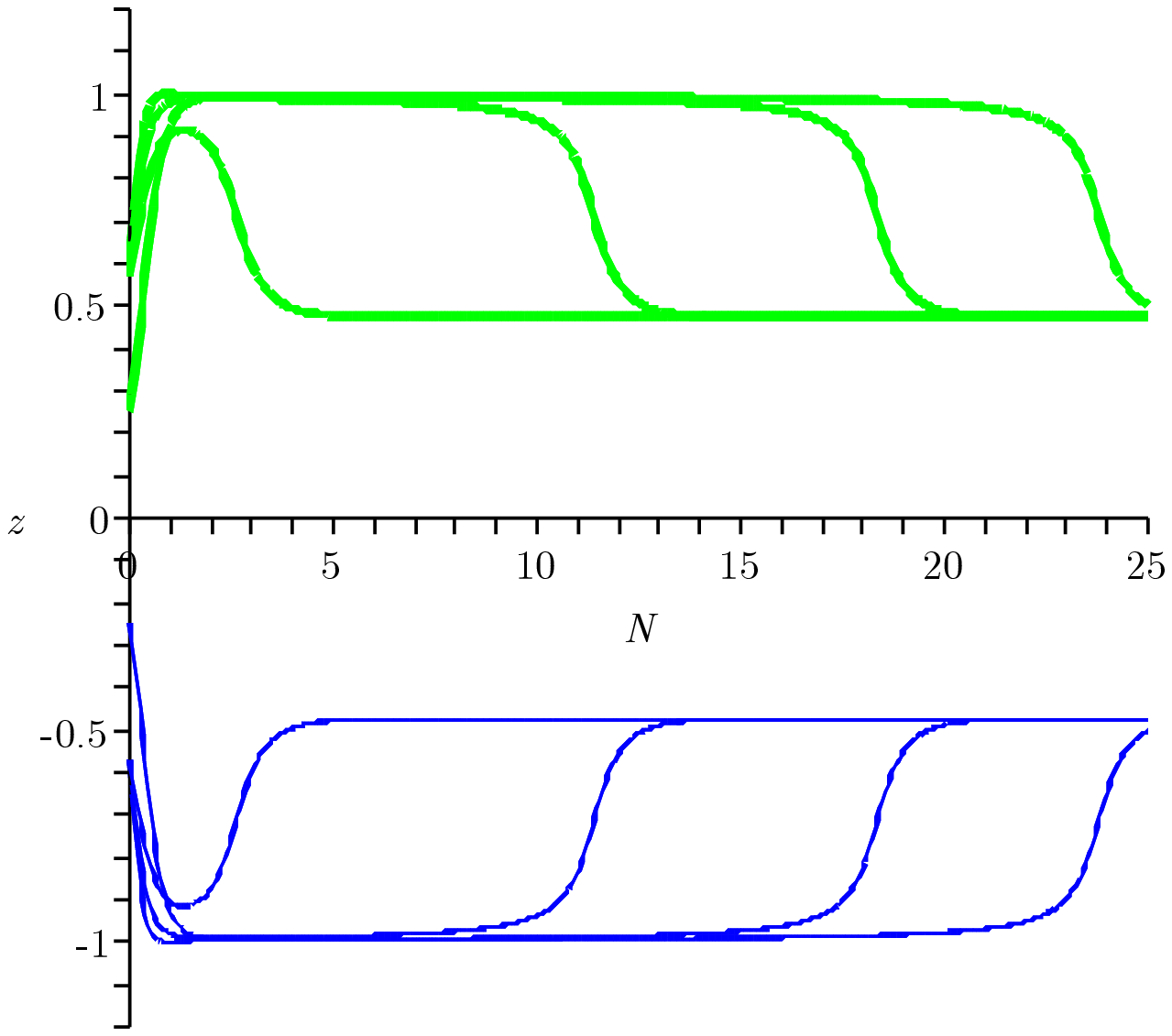}
\caption{(Case I) Evolution of $x, y, z$ versus e-folding number setting $W=1$.} \label{figcase1_xyzN}
\end{center}
\end{figure}

\begin{figure}
\begin{center}
\includegraphics[width=2.7in]{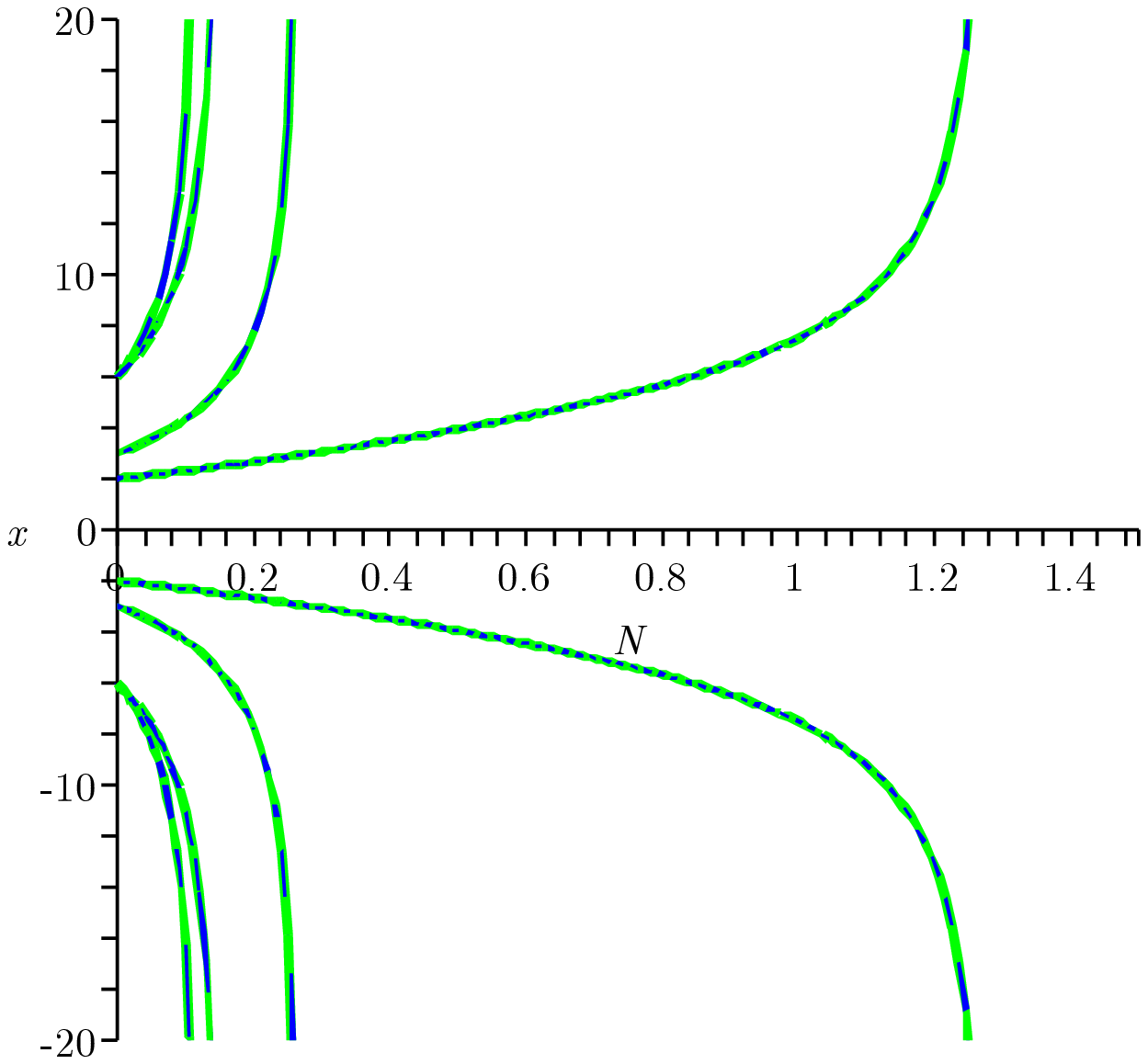}
\end{center}
\begin{center}
\includegraphics[width=2.7in]{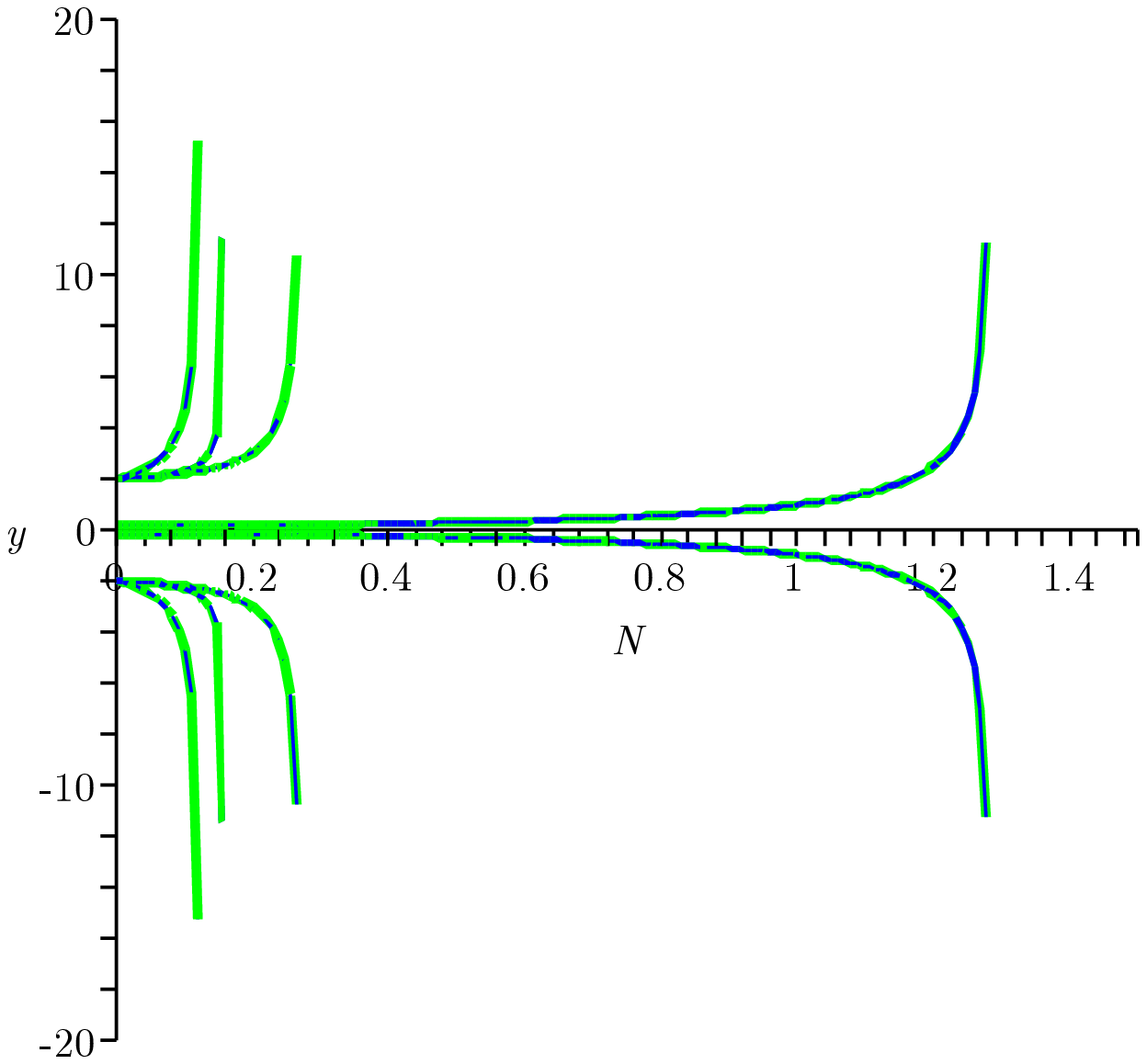}
\end{center}
\begin{center}
\includegraphics[width=2.7in]{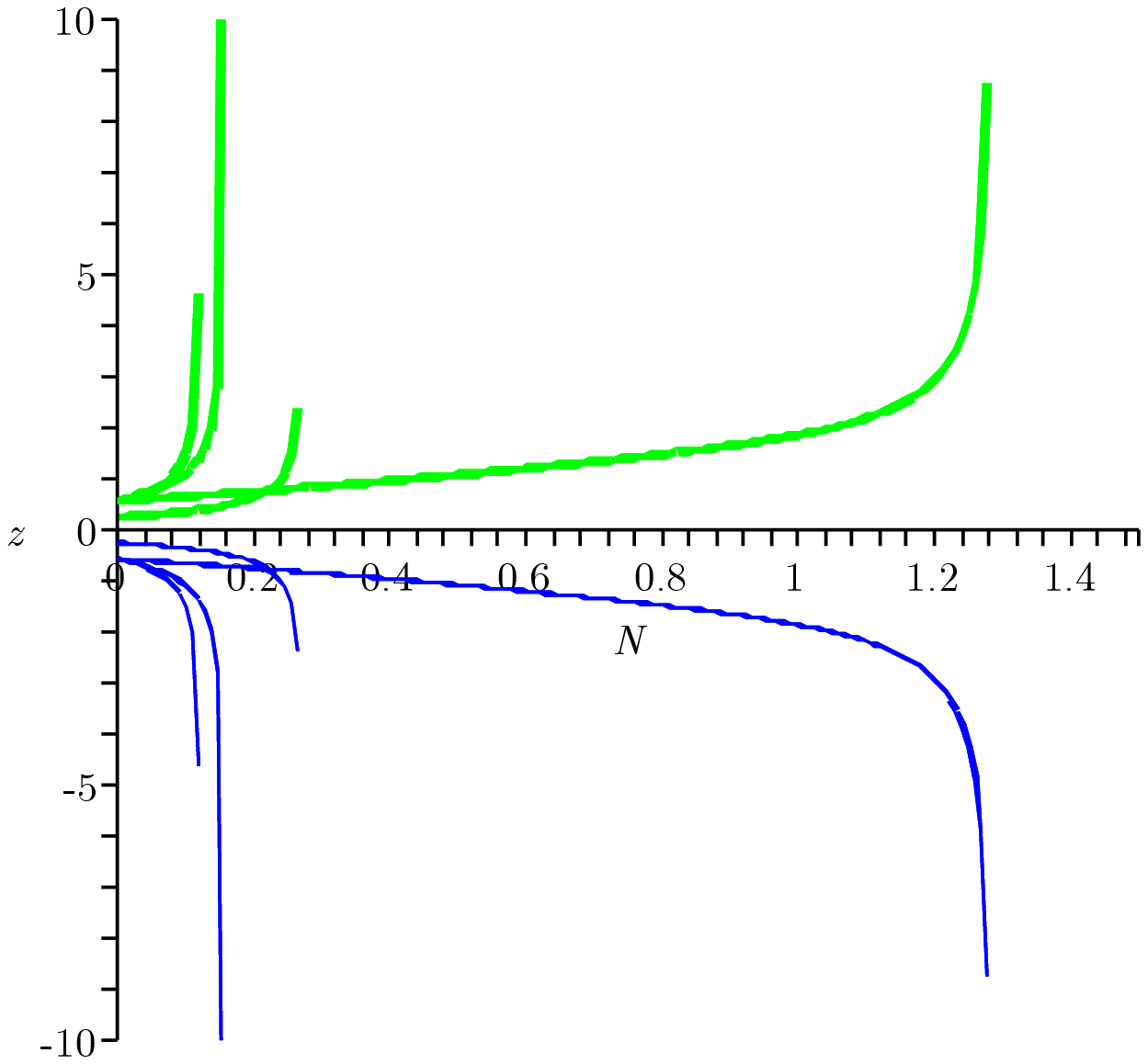}
\caption{(Case I) Evolution of $x, y, z$ versus e-folding number setting $W = 0.95$. All solutions diverge from the
origin} \label{figcase1w<1}
\end{center}
\end{figure}

Figs. \ref{figcase1_xyz} and \ref{figcase1_xy} show the numerical solutions in phase space. For $W=1$ case, the numerical constants are given as
$M_p =1, m=1, \lambda =1$ and $w = 0$ (dust case). Other parameters are $\alpha = 4, \beta=2$ and $A = 2\sqrt{2}$. As expected the (five) fixed
points all lie along the curve $y^2=3x^2$. We also plot the evolution of each parameter $(x,y,z)$ as a function of the e-folding number in Fig.
\ref{figcase1_xyzN} where each of the coordinates tends to its critical value. As expected the phase space dynamics are $\mathbb{Z}_2$ symmetric
about the origin. Note that in the case of $y(N)$ one can keep $y$ suppressed for a few e-foldings with enough tuning, before eventually it
evolved towards the points $\pm \sqrt{3}$ at late times. The full numerical solution of the case $W<1$ is illustrated in Fig. \ref{figcase1w<1}
where $W = 0.95,$ which uniquely fixes the critical points to be $x=\pm \sqrt{20}, y=0, z=1$. As one can see from the resulting plot, this is an
unstable node because the general behaviour is divergent. Note that $x \to \infty$ in this regime effectively solves all the dynamical equations
trivially.


\newpage

\subsection{Case II}
Analogous to the first case, let us now consider another branch of solutions where this time the tension of the brane is taken to be constant. This
dramatically alters the relativistic rolling of the scalar field since the $\gamma$ factor is no longer warped. Initially let us consider the ansatz
\begin{equation}
\tilde V(\phi) = \frac{m^2 \phi^2}{2}\,,   \hspace{1cm} T(\phi)=T\,,   \hspace{1cm} W(\phi) = \frac{\phi^4}{\lambda^4}\,,
\end{equation}
which implies that
\begin{equation}
\mu_1 = \left(\frac{4\sqrt{2 T^3}M_p}{\lambda^4 m^3}\right)\frac{z^2\gamma}{x^2}\,, \hspace{1cm} \mu_2 =0\,, \hspace{1cm} \mu_3 = \frac{2z
\mu_1}{\gamma^2 x} \,,
\end{equation}
and the corresponding field equations become
\begin{eqnarray}
x' &=& -\frac{\alpha \gamma y z^5}{x^4}-\frac{y^2}{2x}-\frac{x H'}{H}\,, \nonumber \\
y'&=& -3y\left(1-\frac{y^2}{6x^2} \right) \left(1+\frac{\alpha \gamma z^5}{2x^3 y} \right) -\frac{6 \alpha z^3}{\gamma x} - \frac{yH'}{H}\,, \nonumber \\
z' &=&\frac{\alpha \gamma z^4}{2x^3}-\frac{z H'}{H}\,,
\end{eqnarray}
where we have defined $\alpha$ as the constant pre-factor in the definition of $\mu_1$.

As before we separate the solution space into two - first finding solutions to $z=0$ and then solutions to $H'/H=\alpha \gamma z^3/(2x^3)$. In the
first case is is straightforward to see that there are the usual  fixed point solutions at $(0,0,0)$ and $(1,\sqrt{3},0)$ (with their respective partner solutions)
respectively coming from the usual condition that $y^2=3x^2$.
The secondary branch of solutions also admit fixed points when $y=0$, however the condition on $z$ is that $z=0, -4x^2$. Since we want real solutions
we are forced to set $z=0$ as a secondary constraint. This forces $W$ to diverge and therefore in the limit that $z \to 0$ we find that $x^2 \to \pm 1$
which is a unique solution. Again the density parameter vanishes identically in this limit as one would expect. The remaining solutions are actually
extremely difficult to solve analytically as they correspond to high order polynomials. As a result we are forced to sketch their behaviour numerically.

Phenomenologically we see that the ansatz presented above is a special class of the more general solution
\begin{equation}
T(\phi)= T\,, \hspace{0.5cm} \tilde V(\phi) = \frac{m^\beta \phi^\beta}{\beta}\,, \hspace{0.5cm}
W(\phi)=\frac{\phi^{\alpha}}{{\lambda^\alpha}}\,,
\end{equation}
which has the parameterisation constraints
\begin{equation}
\mu_1 = A \left(\frac{z\gamma^{1/2}}{x} \right)^{(2+\beta)/(\alpha-\beta)}\,, \hspace{0.5cm} \mu_2=0\,, \hspace{0.5cm} \mu_3 = B
\gamma^{(2-4\alpha+5\beta)/2(\alpha-\beta)} \left(\frac{z}{x} \right)^{(2+\alpha)/(\alpha-\beta)}\,,
\end{equation}
where $A, B$ are both constants.
One can see from the dynamical equations that fixed points with $z=0$ can only occur when the following condition is met
\begin{equation}
\frac{2(1+\alpha)-\beta}{\alpha-\beta} > 0\,,
\end{equation}
which is trivially satisfied for cases where $\alpha>\beta$ (which we assume as an initial constraint).

More generically we see that, provided $\alpha>-2$, we recover the usual fixed point equation $y^2=3x^2$. However we need to be careful here because if
this condition is satisfied then $W$ becomes undefined. Since this is the overall pre-factor multiplying the DBI action, the action is undefined in this limit
and it should therefore correspond to a point of instability in the phase space.
In the limit where $\alpha= -2$, which implies that $\beta > -2$, the fixed point solution
now lives on the zeros of the polynomial
\begin{equation}
3 x^4 B \gamma^{-5(2+\beta)/2(1+\beta)} + 3 x^2 y - y^3 =0\,,
\end{equation}
which can be used to fix $x=x(y)$ or vice-versa.
This solution is actually indicative of a more general branch of physical solutions where we take $\beta > 2(1+\alpha)$. The resulting fixed point
equation (provided $\alpha \ne 2$) is trivially calculated to be $y^2 = 3x^2$ as before, but now we see that $W$ vanishes identically.
In turn this means that the kinetic terms also vanish and the solution is dominated solely by the potential interaction.
One could imagine a situation such as this occurring in the condensation of an open string tachyon mode on a non-BPS brane, where the vanishing of $W$
indicates that we are living in the closed string vacuum.
For dynamic solutions it seems reasonable to consider this particular case as the late time attractor for the solution $z \to 0$.

\begin{figure}
\begin{center}
\includegraphics[width=3.0in]{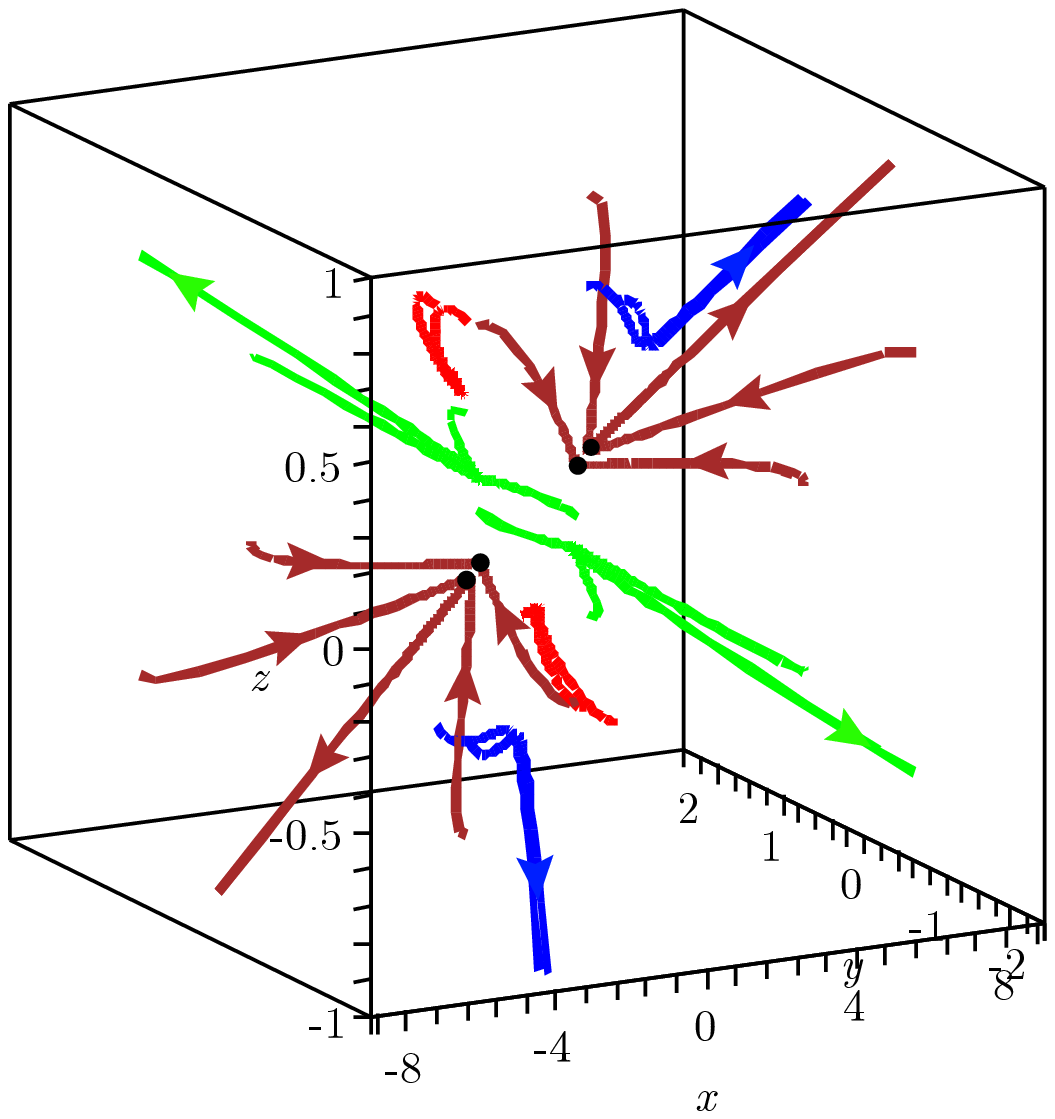}
\caption{(Case II) 3-D $xyz$ phase space trajectories for $T(\phi) = T, \tilde V(\phi) = m^2 \phi^2 /2$ and $W(\phi) = \phi^4/\lambda^4$. We
have set here, $M_p =1, T=1, m=1, \lambda =1$ and $w = 0$ (dust case). } \label{figcase2_xyz}
\end{center}
\vspace{0.5cm}
\begin{center}
\includegraphics[width=2.7in]{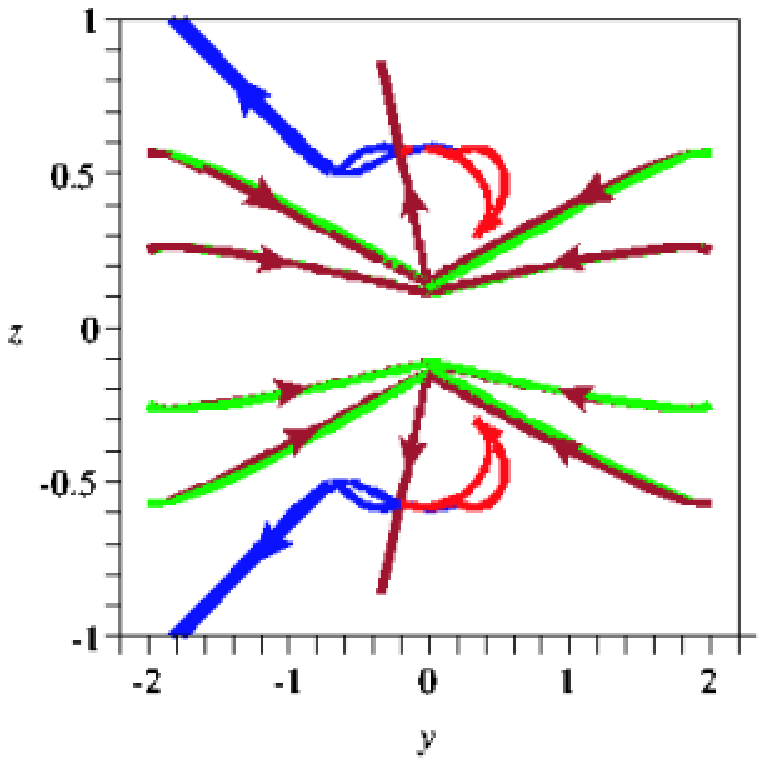}
\caption{(Case II) Trajectory slice through the $yz$ plane.} \label{figcase2_yz}
\end{center}
\end{figure}

\begin{figure}
\begin{center}
\includegraphics[width=2.7in]{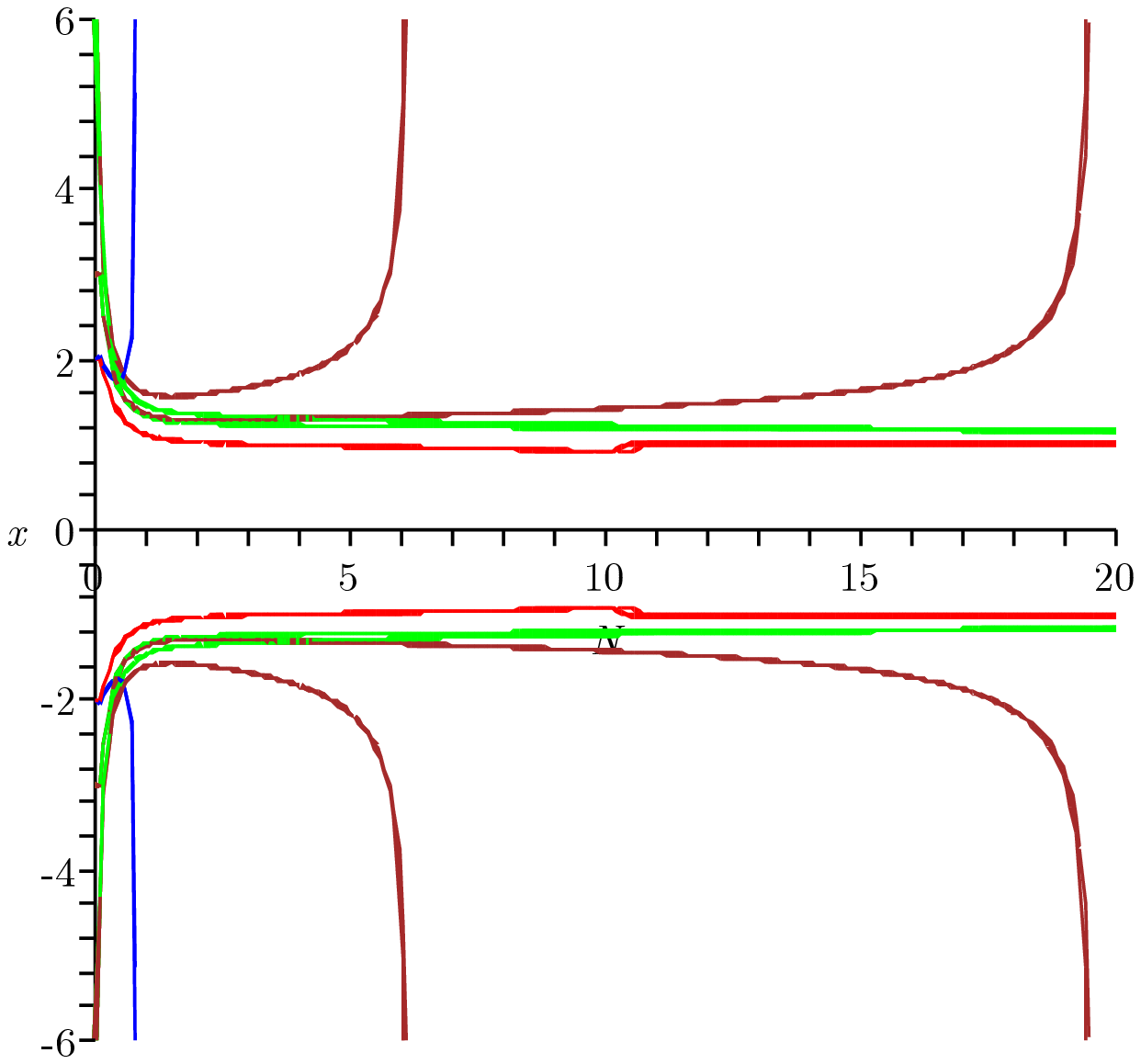}
\end{center}
\begin{center}
\includegraphics[width=2.7in]{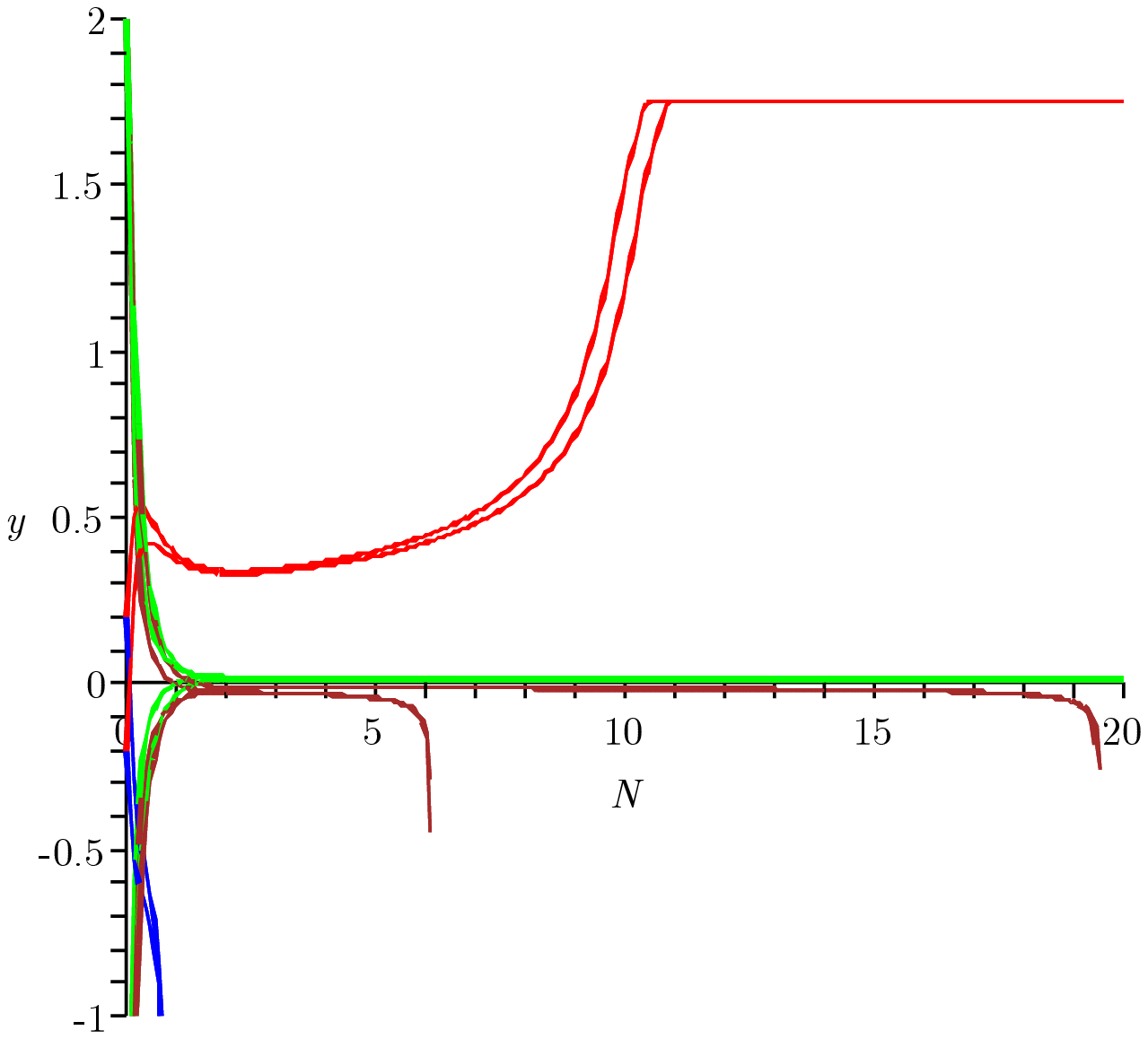}
\end{center}
\begin{center}
\includegraphics[width=2.7in]{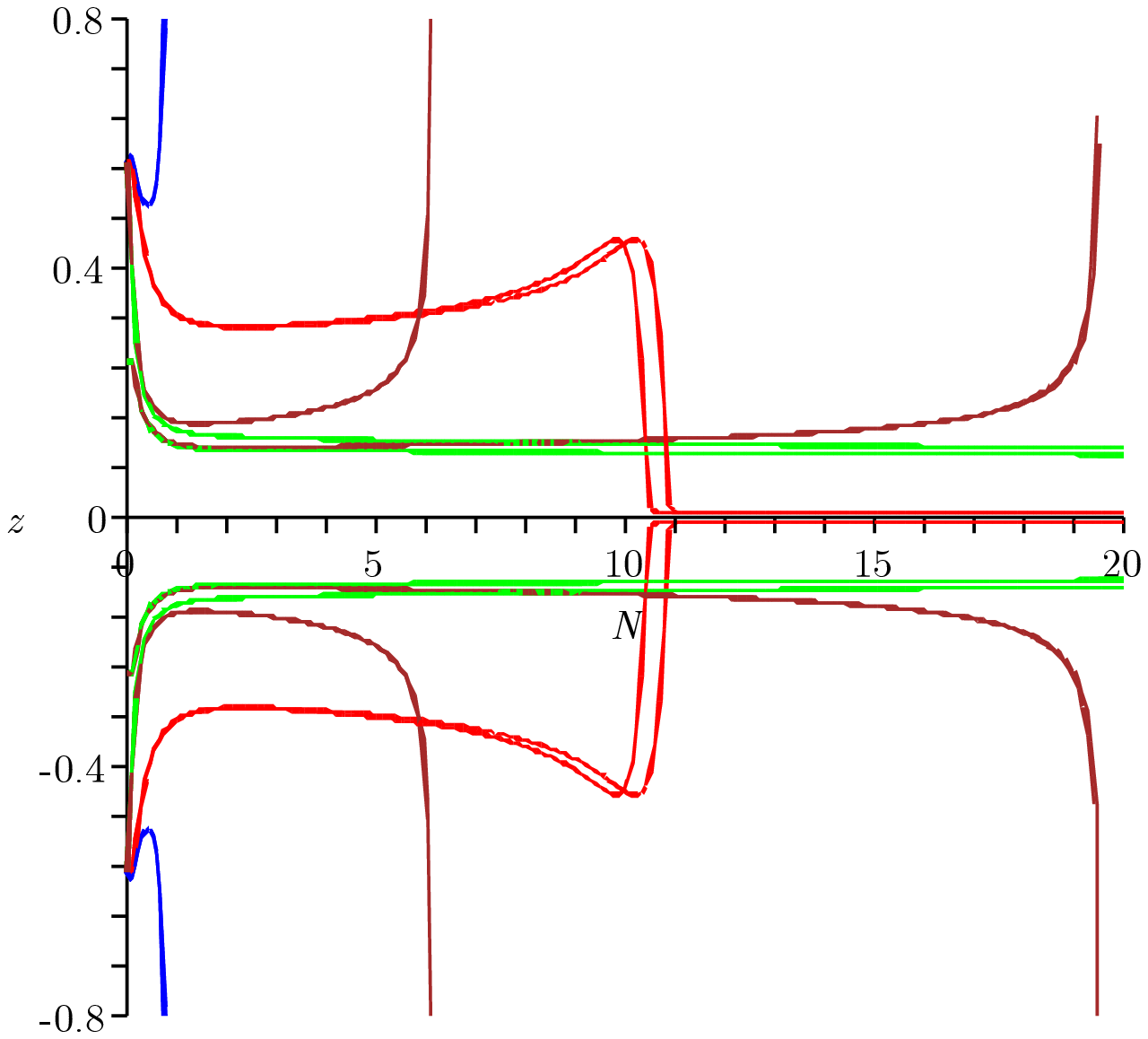}
\caption{(Case II) Evolution of $x, y, z$ versus e-folding number.} \label{figcase2_xyzN}
\end{center}
\end{figure}

The second sub-set of solutions is again complicated, but again we can analytically understand the plane at $y=0$, which gives us the fixed point solutions
\begin{equation}
x = \pm \sqrt{-\frac{\beta z^2}{\alpha}}\,, \hspace{0.5cm} y=0\,, \hspace{0.5cm} z = \pm \left(1-\frac{\beta}{\alpha}\left(1-\left\lbrack
-\frac{\alpha T}{\lambda^{\alpha} m^{\beta}} \right\rbrack^{\alpha/(\alpha-\beta)}\right) \right)^{-1/2}\,.
\end{equation}
Clearly for the solution to be real we require that $\alpha, \beta$ have opposite signs. This satisfies our primary constraint, therefore is
a physical possibility. Moreover in the limit where we set $\beta = -\alpha$, we find that $\Omega = 0$ which is again the dust solution.
Illustrations of  numerical solutions for the case II are in Figs. \ref{figcase2_xyz}, \ref{figcase2_yz} and \ref{figcase2_xyzN}. Constants are
set as $M_p =1, T=1, m=1, \lambda =1$ and $w = 0$ (dust case). Other parameters are $\alpha = 4, \beta=2$. From the numerical analysis one sees
that there are six saddle nodes, only two attractors and one repulsive point which is the origin $(0,0,0)$ as expected.
The dynamical trajectories are particularly interesting due to their apparent lack of monotonicity as a function of e-fold number. The $z$ term in particular
appears to have a large variation in trajectory, diverging in some instances whilst rapidly reaching zero in other instances. Conversely the $y$ variable
displays very uniform (physical) trajectory behaviour, with several curves almost on top of one another at $y=0$ and the remainder smoothly driven to the
(unstable) critical point $y_c \sim 1.8$ in the example given.


\newpage

\subsection{Case III}
Let us now consider a new case where only $W=W(\phi)$, with all the other terms being constant.
We will take $W = \phi^\alpha/\lambda^\alpha$ for generality - which in turn should impose a constraint on the allowed values of $\alpha$.
In this case we see that
\begin{equation}
\mu_1 = 0 \hspace{1cm} \mu_2 =0 \hspace{1cm} \mu_3 = \alpha A \left(\frac{z}{x} \right)^{(\alpha+2)/\alpha} \frac{1}{\gamma^{(2\alpha-1)/\alpha}}
\end{equation}
where $A$ is a function of the constant parameters $A = M_p /\lambda (T/\tilde V)^{(\alpha+2)/2\alpha}$. Because only $\mu_3$ is non-zero the resulting
dynamical expressions are considerably easy to work with
\begin{eqnarray}
x'&=& - \frac{y^2}{2x}- \frac{xH'}{H} \nonumber \\
y'&=& -3y \left(1-\frac{y^2}{6x^2} \right)-3\alpha A z^{(\alpha+2)/\alpha} x^{(\alpha-2)/\alpha}-\frac{yH'}{H} \nonumber \\
z'&=& -\frac{zH'}{H}
\end{eqnarray}
Considering the slice again through $z=0$, we see that the  solutions split into two types depending upon the integer $\alpha$. We recover the usual $y^2=3x^2$ curve
only when $\alpha >0$ or when $\alpha <-2$. If $\alpha=-2$ then the corresponding polynomial equation becomes
\begin{equation}
y \gamma^{9/2} = 2 A x^2
\end{equation}
which is difficult to solve analytically due to the dependence of $\gamma$ on both $x,y$. This expression
does not admit anything but the trivial solution if we set $y$ to zero\footnote{By trivial we mean the point $(0,0,0)$}.
Again we see that there is a potential problem here since the
potential $W$ goes like $1/z^2$, and is therefore divergent in this limit. Solutions to this expression are possible, but complicated. Interestingly there
does exist a solution curve given by
\begin{equation}
y^2 = a x_c^2, \hspace{1cm} x_c = \frac{81}{2A} \frac{\sqrt{3a}}{(9-3a)^{9/4}}
\end{equation}
where the parameter $a$ factor must satisfy $ 0 \le a < 3$ for this solution to be physical. Since $a$ need not be integer, there are essentially a continuum
of curves giving rise to fixed points in this theory.

The secondary branch of solutions again admit fixed point behaviour for $y=0$, however things are more complicated since the fixed points are now
obtained by solving more non-linear expressions. There are two cases of immediate interest however. Firstly if we have $\alpha=2$ then we see that
$z^2 = -1/(2A)$ which is only real when $A$ is negative. Since we have chosen our parameterisation such that this quantity is positive,
this particular branch of solutions is ruled out. Interestingly when $\alpha=-2$ there is a unique fixed point located at
\begin{equation}\label{eq:case3soln}
x = \pm \frac{1}{2A}, \hspace{0.5cm} y=0, \hspace{0.5cm} z = \pm \frac{1}{\sqrt{T/\tilde V-1}}\sqrt{\frac{1}{2A}-1}.
\end{equation}
which corresponds to a positive definite equation of state parameter
\begin{equation}
\Omega = \frac{2T^2 A(A-1)+\tilde V^2(T/\tilde V -1)}{AT\tilde V(T/\tilde V-1)(2A-1)}.
\end{equation}
Note that we must require $T > \tilde V$ for this solution to be non-singular, which means (again) that the tension term dominates the
energetics of the theory. What is also obvious is that demanding $A=1/2$ leads to a novel fixed point at $( \pm 1,0,0)$ regardless of the ratio
$T/ \tilde V$. Using the definition of $A$ this fixes $\lambda = 2 M_p$ and therefore $W$ is vanishingly small unless the scalar takes is trans-Planckian.
This is manifest in a divergence in the equation of state parameter and is therefore unphysical. Therefore we must ensure that $A < 1/2$
implying that $\lambda > 2 M_p$. Since this is the largest scale in our theory, one again expects this to be unphysical.

\begin{figure}
\begin{center}
\includegraphics[width=3.0in]{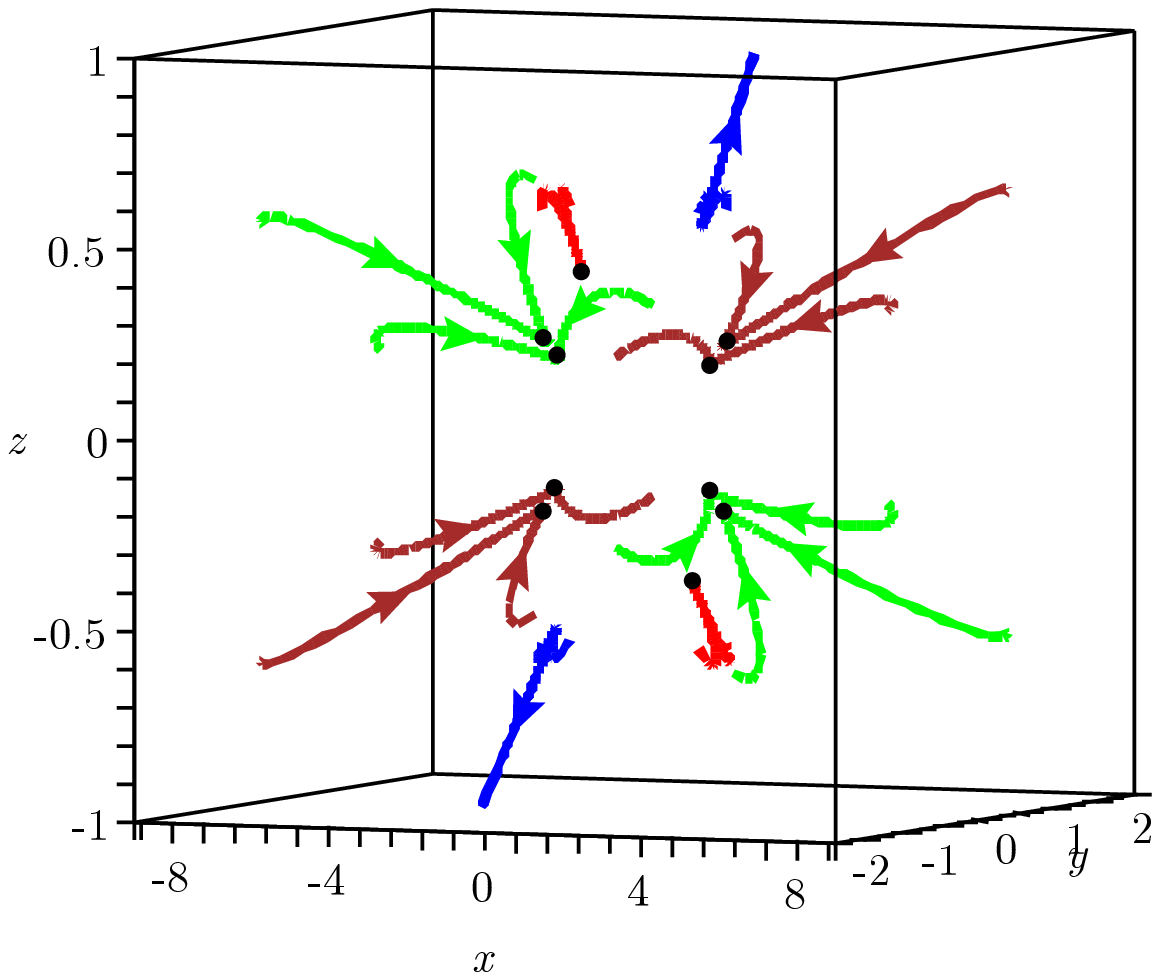}
\caption{(Case III) 3-D $xyz$ phase space trajectories for $T(\phi) = T, \tilde V(\phi) = V$ and $W(\phi) = \phi^{\alpha}/\lambda^{\alpha}$. We
have set here, $M_p =1, T=V=1, m=1, \lambda =1$ and $w = 0$ (dust case), $\alpha = 1$. Green lines approach an attractor} \label{figcase3_xyz}
\end{center}
\vspace{0.5cm}
\begin{center}
\includegraphics[width=2.7in]{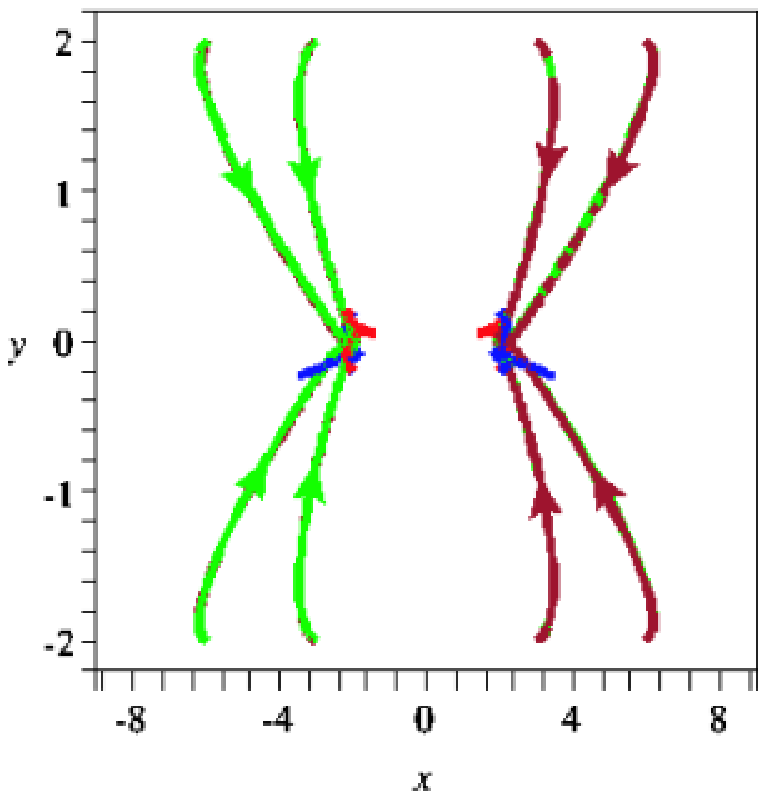} \caption{(Case III) Phase space trajectories in $xy$ plane.} \label{figcase3_xy}
\end{center}
\end{figure}

\begin{figure}
\begin{center}
\includegraphics[width=2.7in]{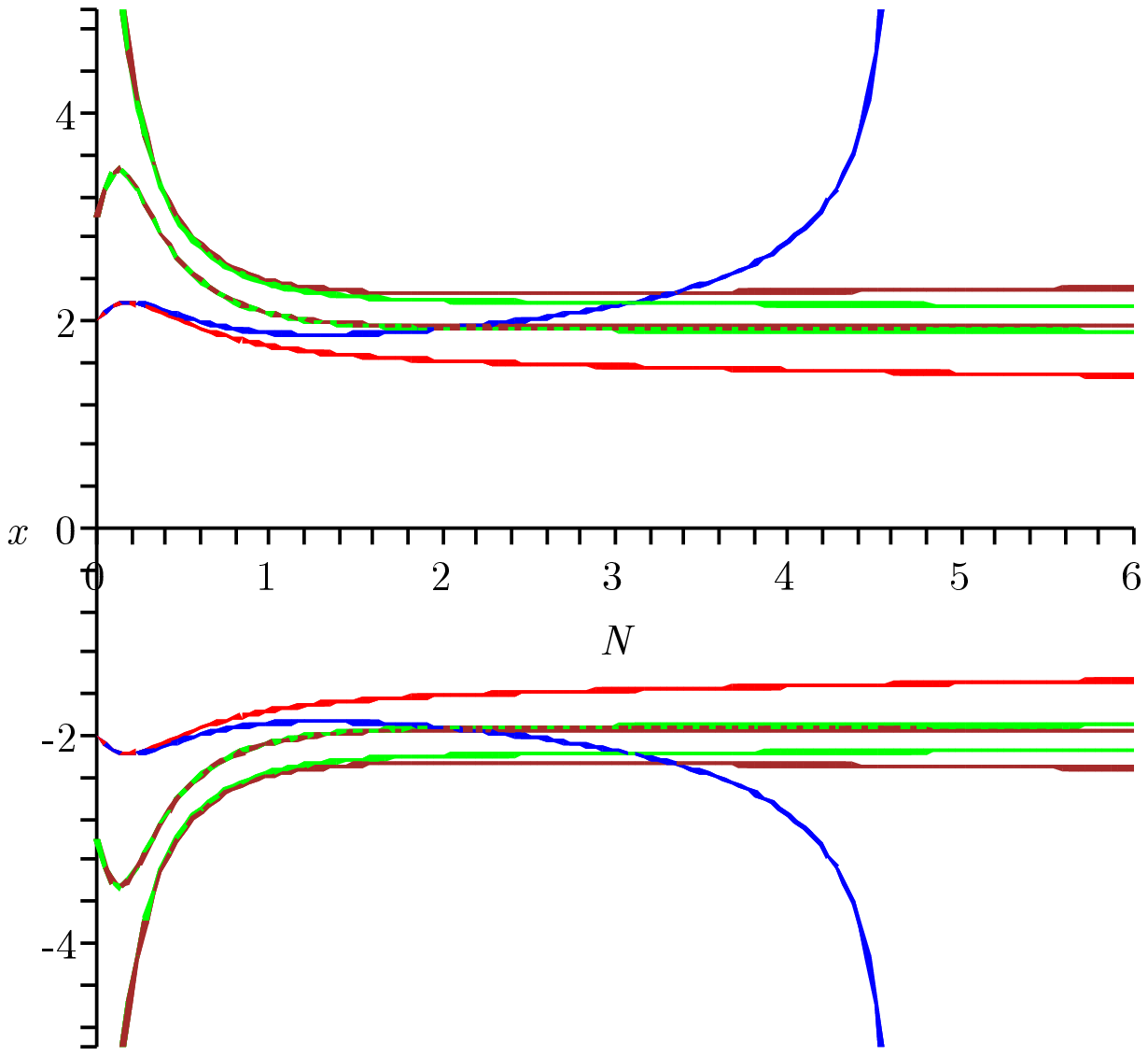}
\end{center}
\begin{center}
\includegraphics[width=2.7in]{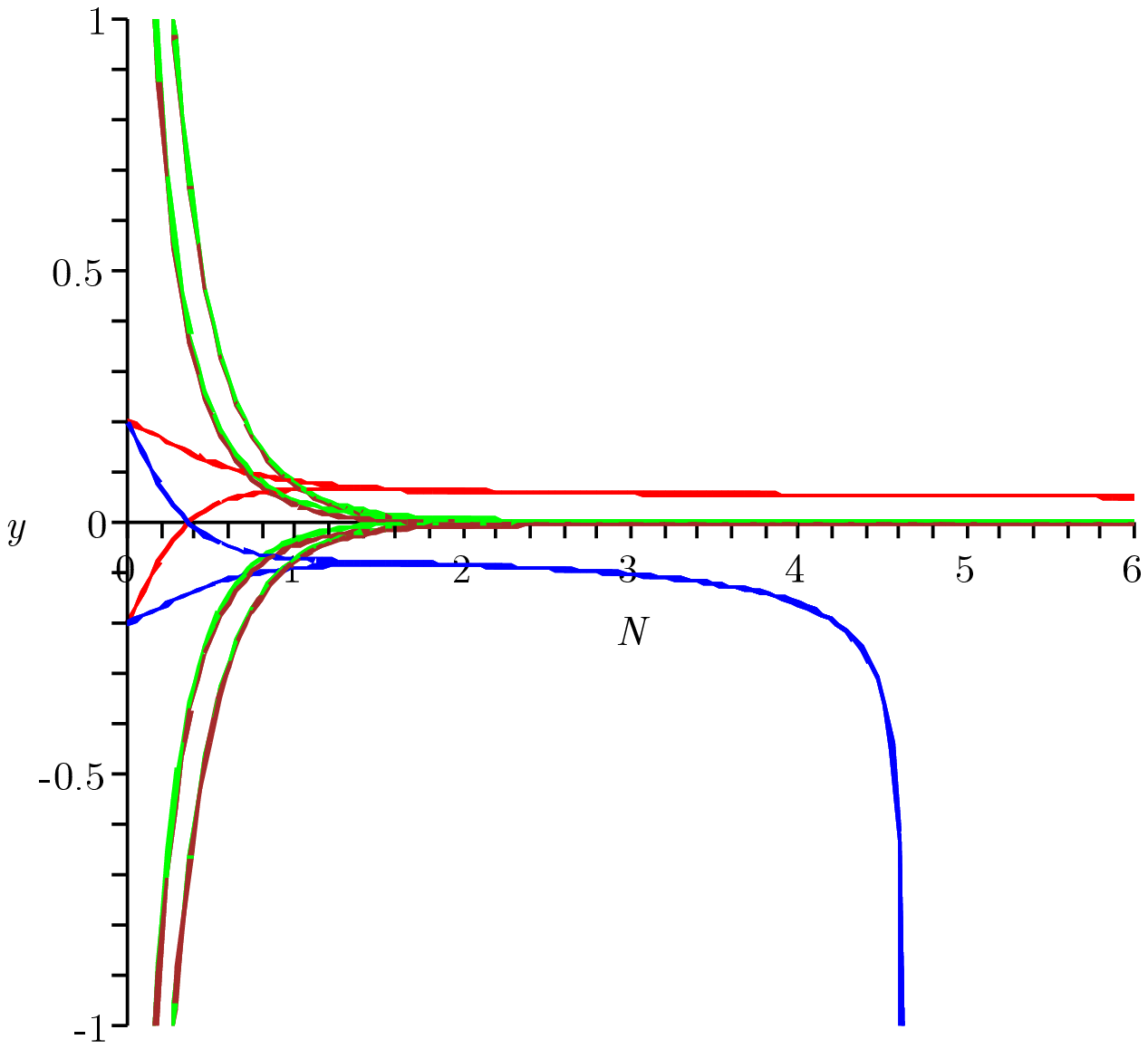}
\end{center}
\begin{center}
\includegraphics[width=2.7in]{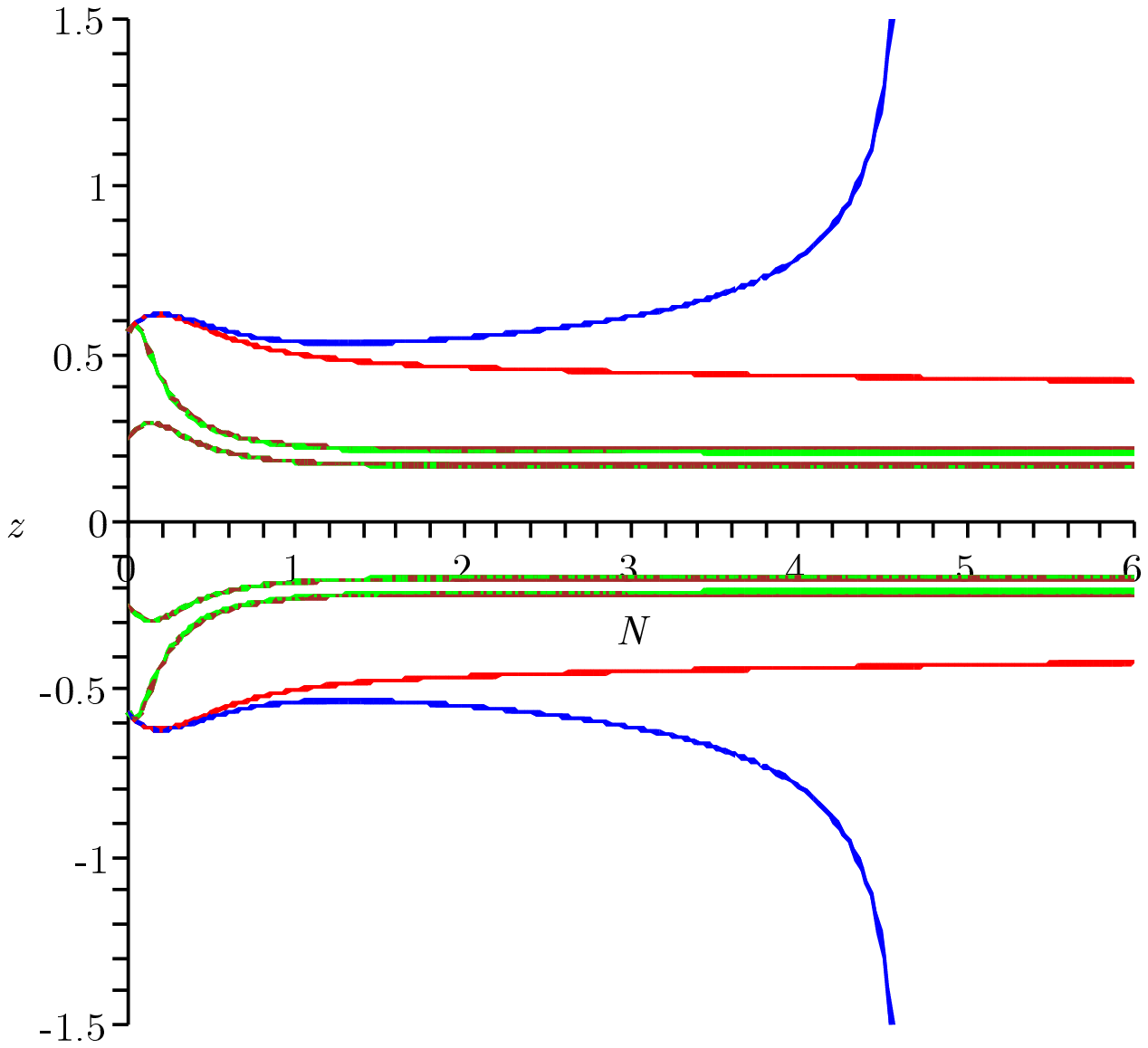}
\caption{(Case III) Evolution of $x, y, z$ versus e-folding number.} \label{figcase3_xyzN}
\end{center}
\end{figure}

The more general solutions can be found numerically and correspond to $x_0^2 = 1+z_0^2 (T/\tilde V-1)$ where $z_0^2$ are the characteristic
solutions to the non-linear equation
\begin{equation}
1+\alpha A z^{(2+\alpha)/2} \left(1+z^2(T/\tilde V -1) \right)^{(\alpha-2)/2\alpha}=0.
\end{equation}
In this more general case we can set $T=\tilde V$ without the solution diverging, and we therefore find the corresponding fixed point solution
is thus given by
\begin{equation}
x = \pm 1, \hspace{0.5cm} y=0, \hspace{0.5cm} z = \left(-\frac{1}{A\alpha} \right)^{2/(2+\alpha)}
\end{equation}
which implies that $\alpha$ is negative. Moreover we see that $\Omega$ is again zero here for all physical values of $\alpha$, although there is no
additional constraint upon the magnitude of $A$.
Now, we see numerical solutions in Figs. \ref{figcase3_xyz},  \ref{figcase3_xy} and \ref{figcase3_xyzN}. Constants are set as $M_p =1, T=1,
\tilde V = 1, \lambda =1$ and $w = 0$ (dust case). Other parameters are $\alpha = 1$ and $A = 1$.



\newpage
\subsection{Case IV}
Following on from the previous class of models, we can find solutions where the scalar potential is now constant, using the ansatz.
\begin{equation}
\tilde V = V, \hspace{0.5cm} T(\phi) = \left(\frac{\phi}{\lambda}\right)^{\alpha}, \hspace{0.5cm} W(\phi) = \left(\frac{\phi}{\delta} \right)^{\beta}
\end{equation}
where $\lambda, \delta$ are terms of the requisite dimensionality. From this expression we see that $\mu_1$ is identically zero. It will be
convenient to define the following function $Q= V \lambda^{\alpha} \delta^{\beta}$ which in turn can be used in the definitions of the remaining
$\mu_i$ functions
\begin{eqnarray}
\mu_2 &=& - \frac{\alpha M_p}{\lambda^{\alpha/2} V} \left(\frac{Q x^2}{\gamma z^2} \right)^{n_1} \\
\mu_3 &=& \frac{\beta M_p \delta^{\beta/2}}{\gamma^{4/2}} \left(-\frac{\mu_2 \lambda^{\alpha/2}V^{3/2}}{\alpha M_p} \right)^{-n_2} \nonumber \\
n_1 &=& \frac{3\alpha-2}{2(\alpha+\beta)}, \hspace{1cm} n_2 = \frac{1+\beta}{3\alpha-2} \nonumber
\end{eqnarray}
and now the dynamical equations simplify to become
\begin{eqnarray}
x' &=& -\frac{\mu_2 y z^3}{2 x^2}-\frac{y^2}{2x}-\frac{x H'}{H} \nonumber \\
y' &=& -3y\left(1-\frac{y^2}{6x^2}\right)\left(1+\frac{z^3 \mu_2}{xy} \right) + 3 \mu_2 \left(\frac{z^{3\alpha+\beta}}{x^{\beta-\alpha}\gamma^{2\beta+\alpha}} \left\lbrack\frac{Q}{\delta}\right\rbrack^{\beta}\right)^{1/(\alpha+\beta)}-3x^2 \mu_3 - \frac{y H'}{H} \nonumber \\
z' &=& -\frac{z H'}{H}.
\end{eqnarray}
The resulting analysis is far more complicated than in the previous cases. Let us again start with the simplest solution slices at $z=0$.
The expressions for $x'$ and $z'$ readily simplify in this instance, however the equation for $y'$ requires us to be more careful.
We see that in order for the $z^3 \mu_2$ term to vanish in this limit we require $(2+3\beta)/(\alpha+\beta) > 0$. The remaining $\mu_2$ term only
vanishes if this condition is tightened to $(2+\beta)/(\alpha+\beta)>0$ and the term coming from $\mu_3$ only vanishes if $(1+\beta)/(\alpha+\beta)>0$. If
these inequalities are reversed, for example, then these terms diverge in the $z \to 0$ limit. If we restrict ourselves to well-behaved solutions such that
$\alpha, \beta$ satisfy the above bounds (either by both $\alpha, \beta \ge 0$ or by $\alpha \ge 0, \beta \le 0$ with $|\beta|>|\alpha|$),
then we obtain the solution curve $y^2=3x^2$ as usual. If the parameters $\alpha, \beta$ do not satisfy at least the minimal bound, then one can only
solve these expressions numerically.

The only other solution branch occurs when $H'/H=0$. This is again a complicated solution, however things simplify somewhat when we slice through $y=0$, but also
tune the solution such that $\alpha=\beta$, which gives us
\begin{equation}
z= \frac{x \delta^{\alpha}}{2\sqrt{Q}} \left(1 \pm \sqrt{1-\frac{4Q(x^2-1)}{x^2 \delta^{2\alpha}}} \right)
\end{equation}
and therefore the fixed point solution in this instance is given by solutions of the polynomial
\begin{equation}
x\sqrt{\frac{Q}{\delta}}+ \left(\frac{\sqrt{Q}}{z}\right)^{1/(2\alpha)}\frac{x^{(1+8\alpha)/(4\alpha)}}{\lambda^{\alpha/2}\delta^{\alpha/2}}=1.
\end{equation}
This can actually be solved exactly when $\alpha=-1$, but numerically for more general $\alpha$. The exact case gives us the following
solution
\begin{eqnarray}
x_0 &=& \frac{Q\delta - 2\sqrt{Q}+\delta^2 \pm \delta \sqrt{F(\lambda, \delta)}}{2\lambda \delta^4} \\
F(\lambda, \delta) &=&Q^2+\delta^2-4 \sqrt{Q^3 \delta}+6Q\delta - 4\sqrt{Q\delta^3}+16 \sqrt{Q^3 \delta^7}-4Q^3 \delta^2  \nonumber \\
&+& 16\sqrt{Q^{5/2}\delta^{5/2}} -24Q^2 \delta^3-4Q \delta^4 + 4\delta^6 \lambda \nonumber
\end{eqnarray}
where $z_0$ is given by the term written above. This is a highly complicated solution, but one sees that in principle there are many fixed points
along the plane $(x_0, 0, z_0)$ depending on the constants $\lambda ,\delta$. One also sees that there is a simple solution when $x=1$, since
this implies that $z_0 = \delta^{\alpha}/\sqrt{Q}$ or $z_0 = 0$, the latter again giving rise to the point $(1,0,0)$ which corresponds to the non-propagating
end point of the brane dynamics.

\begin{figure}
\begin{center}
\includegraphics[width=3.0in]{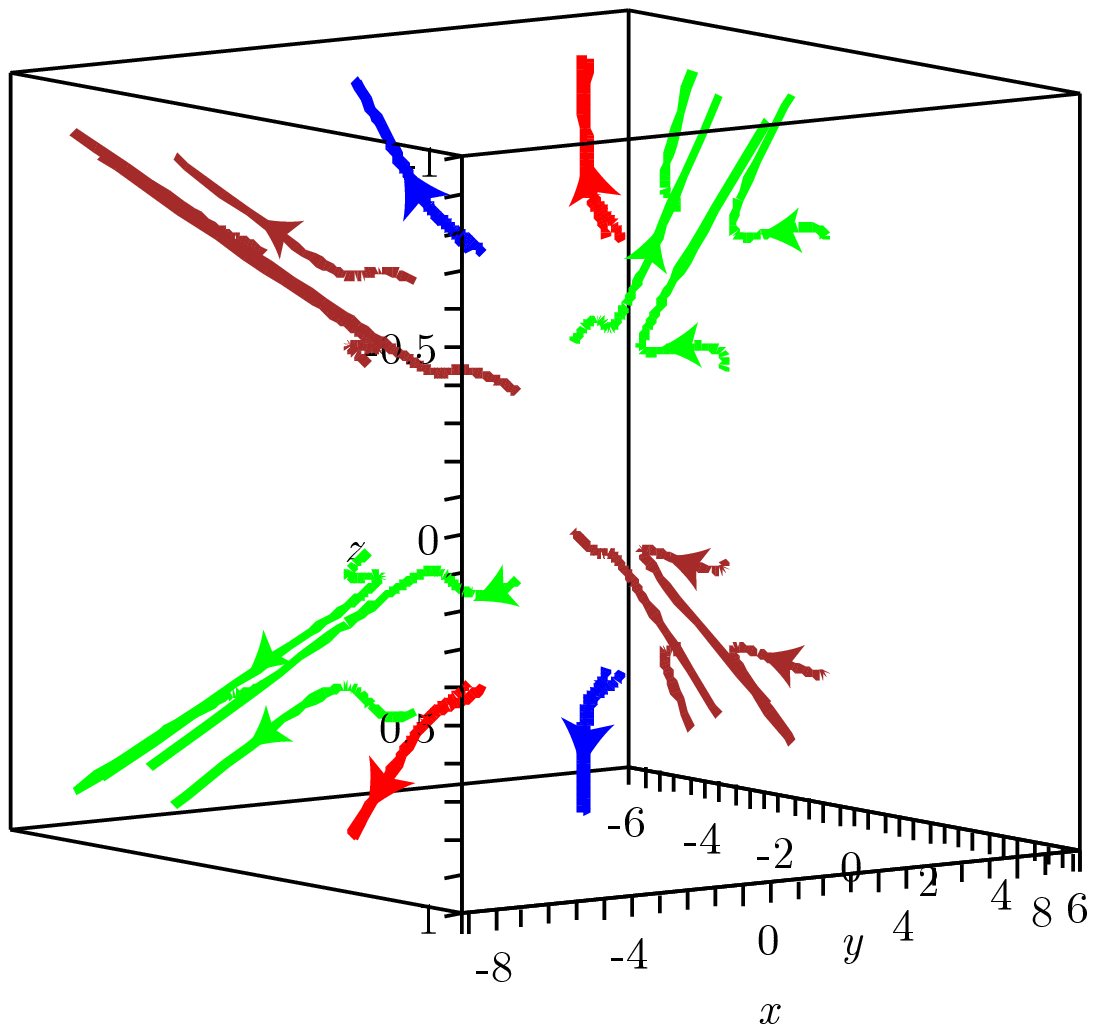}
 \caption{(Case IV) 3-D $xyz$ phase space trajectories for
 $T(\phi) = (\phi/\lambda)^{\alpha}, \tilde V(\phi) = V$ and $W(\phi) = (\phi/\delta)^{\beta} $. Here, $M_p =1, V=1, m=1,
\lambda =1, \alpha=1, \beta=1, \delta=1 $   and $w = 0$ (dust case)} \label{figcase4_xyz}
\end{center}
\vspace{0.5cm}
\begin{center}
\includegraphics[width=2.7in]{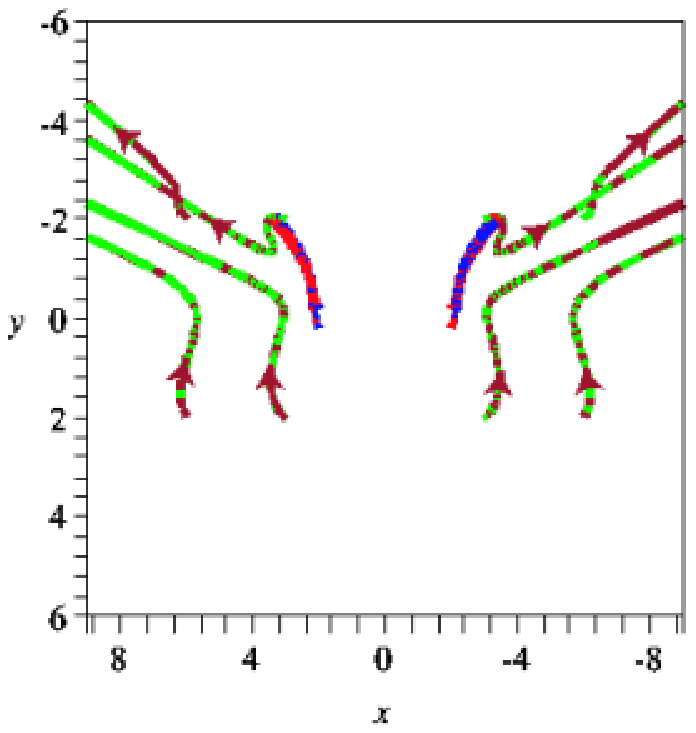} \caption{(Case IV) Phase space
trajectories in $xy$ plane.} \label{figcase4_xy}
\end{center}
\end{figure}

\begin{figure}
\begin{center}
\includegraphics[width=2.7in]{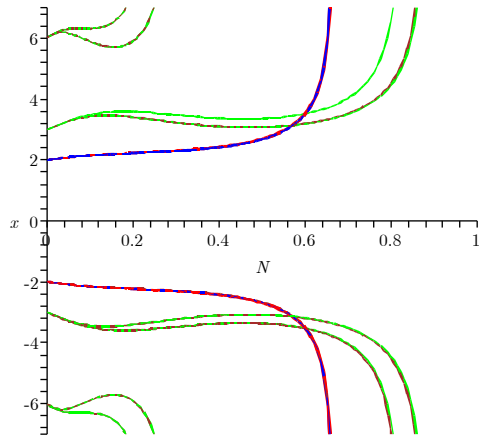}
\end{center}
\begin{center}
\includegraphics[width=2.7in]{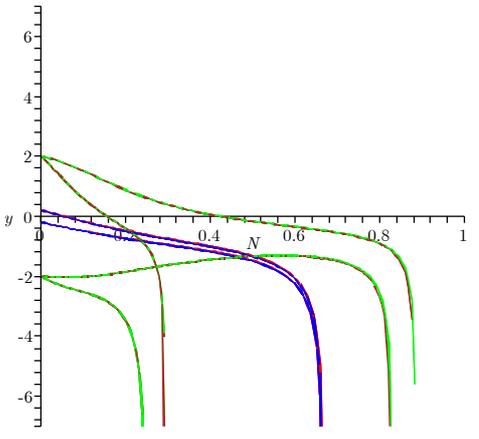}
\end{center}
\begin{center}
\includegraphics[width=2.7in]{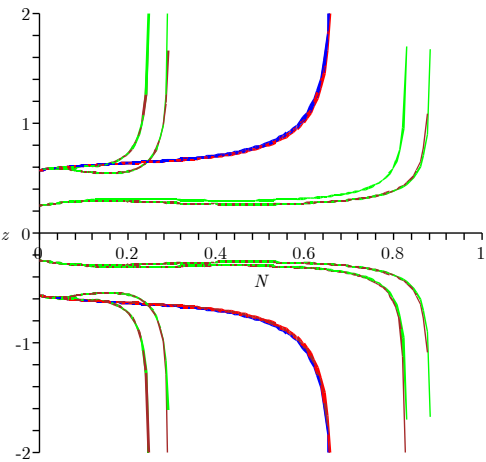}
\caption{(Case IV) Evolution of $x, y, z$ versus e-folding number.} \label{figcase4_xyzN}
\end{center}
\end{figure}

\newpage
\subsection{Case V}
Finally let us comment on perhaps the most general form of the solution one could obtain from this model, namely that corresponding to turning on all the
relevant degrees of freedom. One can therefore see that Cases $I-IV$ are actually slices through the full phase space described in this section.
We will take the following parameterisation for simplicity
\begin{equation}
T = \left(\frac{\phi}{\lambda}\right)^{\alpha}, \hspace{0.5cm} W = \left(\frac{\phi}{\delta} \right)^{\beta}, \hspace{0.5cm} \tilde V =
\frac{m^\xi \phi^\xi}{\xi}.
\end{equation}
In this case we will have all three
$\mu_i$ non zero which complicates the analysis somewhat, and reality again imposes the condition that $\xi>0$. Let us initially search for the
fixed points around $z=0$. The primary constraint equation for this becomes
\begin{equation}
\frac{\alpha-\xi+2(1-\beta)}{(\alpha+\beta-\xi)} > 0
\end{equation}
Let us initially assume that the denominator is positive definite. Going through the same analysis as before yields the usual solution curve $y^2 = 3x^2$ provided that
we tune $\beta >0 $ and $\alpha + \beta > \xi$. However with reference to the action, we see that this situation leads to both $W, T$ diverging
and therefore we should be wary of this part of the solution.
Returning to the constraint equation let us therefore assume that $\xi > \alpha+\beta$ and re-do the analysis. We
then find that the $y^2=3x^2$ is perfectly valid, and moreover the parameters $W,T$ are not divergent provided that the parameters
satisfy $\alpha+\beta-\xi < -(2+\beta)$. Moreover we also see that $\beta$ is bounded from
above such that $\beta < -2/3$ - thus severely restricting the form of the variable phase space.

If we search for solutions along the $y=0$ slicing things are again complicated.
However we can simplify things by identifying $\alpha = \xi$, since we can then solve
explicitly for $x$ via
\begin{equation}
x^2 = 1 + z^2\left(\frac{\xi^2}{\lambda^{\xi}m^{\xi}}-1\right).
\end{equation}
The remaining equation coming from $y'=0$ has several solutions. The simplest being $z^2=0, (\lambda^{-\xi} m^{-\xi}\xi^2-1)^{-1}$
which give rise to the points
\begin{eqnarray}
x_0 &=& \pm \sqrt{2}, \hspace{0.5cm} y_0 =0, \hspace{0.5cm} z_0=\frac{1}{\sqrt{\lambda^{-\xi}m^{-\xi}\xi^2-1}} \nonumber \\
x_0 &=& \pm 1, \hspace{0.8cm} y_0=0, \hspace{0.5cm} z_0 = 0
\end{eqnarray}
however the first of these conditions also requires that $\xi^{2/\xi}>\lambda m$ for the solution to be real. The maximal value of $\xi^{2/\xi}$ is actually
given by $\xi = e^1$ which imposes a tight constraint on the background parameters which can only be satisfied through substantial fine-tuning. Again more
general solutions are only available through numeric methods.





\begin{figure}
\begin{center}
\includegraphics[width=3.0in]{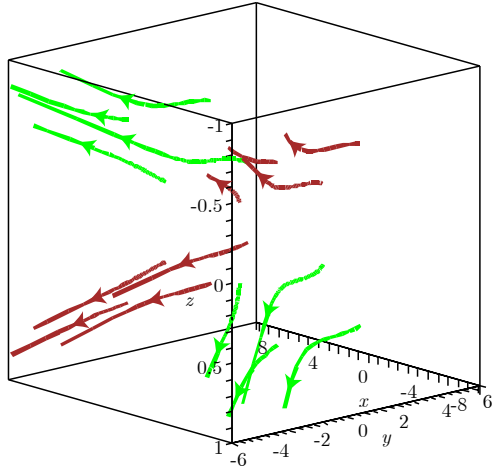}
 \caption{(Case V) 3-D $xyz$ phase space trajectories for
 $T(\phi) = (\phi/\lambda)^{\alpha}, \tilde V(\phi) = (m \phi)^{\xi}/\xi $ and $W(\phi) = (\phi/\delta)^{\beta} $. Here, $M_p =1, V=1, m=1, \lambda =1, \alpha=1, \beta=1, \delta=1, \xi=2 $   and $w = 0$ (dust case)}
\label{figcase4_xyz}
\end{center}
\end{figure}

\begin{figure}
\begin{center}
\includegraphics[width=2.7in]{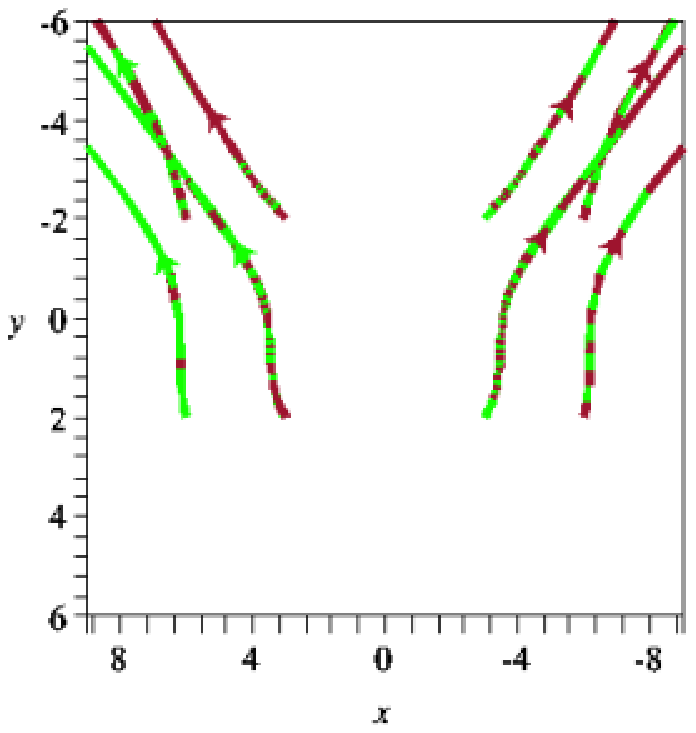} \caption{(Case V) Phase space trajectories in $xy$ plane.} \label{figcase5_xy}
\end{center}
\end{figure}

\begin{figure}
\begin{center}
\includegraphics[width=2.7in]{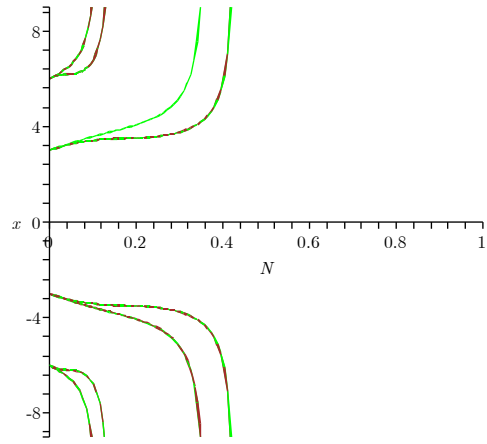}
\end{center}
\begin{center}
\includegraphics[width=2.7in]{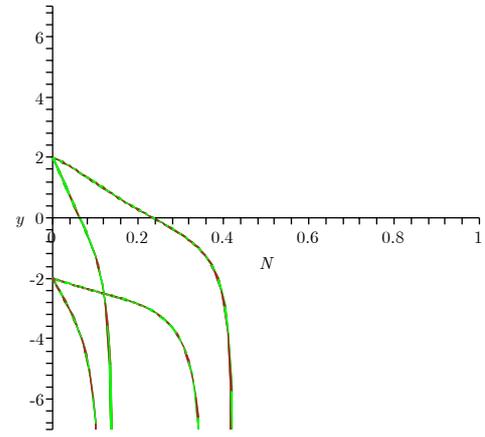}
\end{center}
\begin{center}
\includegraphics[width=2.7in]{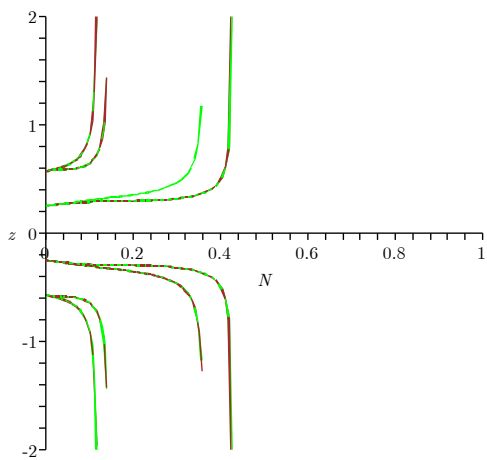}
\caption{(Case V) Evolution of $x, y, z$ versus e-folding number.} \label{figcase5_xyzN}
\end{center}
\end{figure}

\newpage
\section{Perturbations and fixed point stability.}
We now need to evaluate the stability of these fixed point solutions. Clearly one may anticipate that solutions such as $(0,0,0)$ may well be unstable.
We must perturb the field equations about small values; therefore we need
\begin{equation}
x \to x_0 + \delta x, \hspace{1cm} y \to y_0 + \delta y, \hspace{1cm} z \to z_0 + \delta z.
\end{equation}
Now the analysis is more complicated than in standard models due to the complexity of the DBI action and the general (unknown) phase space
dependence of the variables $T, W,\tilde V$. Since $\gamma$ is independent of any particular parameterisation, we can calculate the general result.
\begin{equation}
\gamma \to \gamma \left(1+\frac{\gamma^2 y_0 \delta y}{3 x_0^2}-\frac{\gamma^2 y_0^2 \delta x}{3 x_0^3}+ \ldots \right)
\end{equation}
Using this we can write the perturbation in $H'/H$. In general we can Taylor expand the function $W$ such that we have
$W(x^i + \epsilon^i) \sim W(x^i_0)+\partial_i W \epsilon^i$ and therefore the general result is true
\begin{equation}
\delta \left(\frac{H'}{H} \right) = -y_0 \delta y - \frac{3(1+\omega)}{2}\left(-2 z_0 \delta z-2x_0 \delta x \left\lbrack1-\frac{1}{W\gamma} \right\rbrack
-\frac{x_0^2}{\gamma W}\left\lbrace -\frac{\gamma^2 y_0 \delta y}{3 x_0^2}+\frac{\gamma^2 y_0^2 \delta x}{3 x_0^3} - \frac{\partial_i W \epsilon^i}{W} \right \rbrace \right) \nonumber
\end{equation}
where all terms such as $\gamma, W$ are evaluated on the classical solution and there is a summation over Latin indices.

The general equations even for the linear perturbation, are shown below for Case V - which encompasses all the other solutions in the relevant limit:
\begin{eqnarray}
\delta x' &=& -\frac{yz^3}{2x^2}(\mu_1 + \mu_2)\left(\frac{\delta y}{y} + 3\frac{\delta z}{z}-2\frac{\delta x}{x} \right) - \frac{y z^3}{2x}(\mu_1 \delta \mu_1 + \mu_2 \delta \mu_2) - \frac{y^2}{2x} \left(2\frac{\delta y}{y}-\frac{\delta x}{x} \right) \nonumber \\
&-& \delta x\frac{H_0'}{H_0} - x \delta \left(\frac{H'}{H} \right) \nonumber \\
\delta y' &=& -\frac{3z^3}{x}\left\lbrace\mu_1 \delta \mu_1 + \mu_2 \delta \mu_2 + [\mu_1 + \mu_2]\left(3\frac{\delta z}{z}-\frac{\delta x}{x}-\frac{\delta y}{y} \right)  \right\rbrace - 3x^2 \mu_3 \left(2\frac{\delta x}{x} + \delta \mu_3 \right) \nonumber \\
&+& \left(1+\frac{z^3}{xy}[\mu_1 + \mu_2] \right)\left(\frac{y^3}{x^2}[\frac{\delta y}{y}- \frac{\delta x}{x}]-3 \delta y \right) - \delta y \frac{H_0'}{H_0} - y \delta \left(\frac{H'}{H} \right)\nonumber \\
&=& \frac{3z^3 \mu_2 W}{\gamma x}\left(\delta \mu_2 + 3\frac{\delta z}{z}\left(1-\frac{2\beta}{3n} \right)-\frac{\gamma^2 y \delta y}{3x^2}\left(1+\frac{\beta}{n}+ \frac{\delta x}{x}\left\lbrace \frac{2\beta}{n}-1+\frac{\gamma^2 y^2}{3x^2}\left(1+\frac{\beta}{n}\right) \right\rbrace \right) \right) \nonumber \\
\delta z' &=& \frac{z^2 y \mu_1}{2x}\left(2\frac{\delta z}{z}+\frac{\delta y}{y}-\frac{\delta x}{x} + \delta \mu_1 \right) - \delta z \frac{H_0'}{H_0} - z\delta \left(\frac{H'}{H} \right)
\end{eqnarray}
where we have defined $n=\alpha+\beta - \rho$ for simplicity and also the following terms
\begin{eqnarray}
\delta \mu_1 &=& -\frac{2(\alpha-2\rho)}{n}\frac{\delta z}{z}-\frac{(\alpha-2-\rho)}{2n}\frac{\gamma^2 y \delta y}{3x^2} + \frac{2(\alpha-2-\rho)}{n}\frac{\delta x}{x}\left(1+\frac{\gamma^2 y^2}{12 x^2} \right) \\
\delta \mu_2 &=& -\frac{4(\alpha-1-\rho)}{n}\frac{\delta z}{z}-\frac{(3\alpha-3\rho-2)}{n}\frac{\gamma^2 y \delta y}{6x^2}+\frac{\delta x}{nx}\left(4(\alpha-1-\rho)+\frac{(3\alpha-3\rho-2)\gamma^2 y^2}{6x^2} \right) \nonumber \\
\delta \mu_3 &=& \frac{2(\alpha+2+3\rho+2\beta)}{n}\frac{\delta z}{z}+ \frac{(2\alpha-2-10\rho+\beta)\gamma^2 y \delta y}{6nx^2} + \frac{\delta x}{nx}\frac{(-2\alpha+2+10\rho-\beta)\gamma^2 y^2}{6x^2} \nonumber \\
&-& \frac{\delta x}{x} \frac{(2\alpha+4+6\rho+4\beta)}{n} \nonumber
\end{eqnarray}
We will work through an explicit example to illustrate the formalism, namely
the Case I solutions. Firstly we can calculate the following expression
\begin{equation}
\delta \left(\frac{H'}{H} \right) \sim -y_0 \delta y+\frac{3(1+\omega)}{2}\left(2z_0 \delta z + 2x_0 \delta x\left\lbrack 1-\frac{1}{W\gamma}\right\rbrack +\frac{x_0^2 \gamma y_0}{3 W x_0^2}\left(\delta y - \frac{y \delta x}{x_0}\right) \right)
\end{equation}
which will allow us to calculate the perturbed phase space variables. The perturbed dynamic expressions then take the following form
\begin{eqnarray}
\delta x' &=& \frac{yz^3\mu_1}{2x^2}\left(\frac{\alpha x^2}{\beta W \gamma^2}[\frac{2 \delta x}{x}(1+\frac{\gamma^2 y^2}{6 x^3})-2\frac{\delta z}{z}-\frac{\gamma^2 y \delta y}{3x^2}] \right)-\frac{y^2}{2x}\left(\frac{2\delta y}{y}-\frac{\delta x}{x} \right)\nonumber \\
&-& \frac{yz^3 \mu_1}{2x^2} \left(1-\frac{\alpha x^2}{\beta W \gamma^2}\right)\left(\frac{\delta y}{y}(1-\frac{n \gamma^2 y}{6x^2})+(3-n)\frac{\delta z}{z}+\frac{\delta x}{x}(n-2+\frac{n\gamma^2y^2}{6x^2}) \right) \nonumber \\
&-& \delta x \frac{H_0'}{H_0} - x \delta \left(\frac{H'}{H} \right) \nonumber \\
\delta y' &=& 3z^3 \mu_1\left(1-\frac{y^2}{6x^2} \right)\frac{\alpha x}{\beta W \gamma^2}\left \lbrack2\frac{\delta x}{x}(1+\frac{\gamma^2 y^2}{6x^2})-2\frac{\delta z}{z}-\frac{\gamma^2 y \delta y}{3x^2}\right \rbrack \nonumber \\
&-& \frac{3z^3 \mu_1}{2x}(1-\frac{\alpha x^2}{\beta W \gamma^2})(1-\frac{y^2}{6x^2}) \left((3-n)\frac{\delta z}{z}+\frac{\delta x}{x}(n-1+\frac{n\gamma^2 y^2}{6x^2})-\frac{\delta y}{y}(1+\frac{n\gamma^2y^2}{6x^2}) \right) \nonumber \\
&+& \left(1+\frac{z^3 \mu_1}{xy}[1-\frac{\alpha x^2}{\beta W \gamma^2}] \right)\left \lbrace\frac{y^3}{x^2}[\frac{\delta y}{y}-\frac{\delta x}{x}]-3 \delta y (1-\frac{y^2}{6x^2}) \right \rbrace \nonumber \\
&-& \frac{3xz\alpha \mu_1}{\beta \gamma^2} \left(\frac{\delta x}{x}[1+n+\frac{2\gamma^2 y^2}{3x^2}(1+\frac{n}{4})]+(1-n)\frac{\delta z}{z}-\frac{\delta y}{y}\frac{2\gamma^2 y^2}{3x^2}(1+\frac{n}{4}) \right) \nonumber \\
&-& y \delta \left(\frac{H'}{H} \right) - \delta y \frac{H_0'}{H_0} \nonumber \\
\delta z' &=& \frac{z^2 y \mu_1}{2x}\left((2-n)\frac{\delta z}{z}+\frac{\delta x}{x}[n-1+\frac{n \gamma^2 y^2}{6x^2}]+\frac{\delta y}{y}[1-\frac{n\gamma^2y^2}{6x^2}] \right)  -\delta z\frac{H_0'}{H_0}- z\delta \left(\frac{H'}{H} \right)
\end{eqnarray}
where the notation $H_0'/H_0$ implies that we take this function evaluated at the critical points, and we have defined $n= (\alpha-\beta-2)/(\alpha-\beta)$ for simplicity.
Note that these are the leading order solutions only, and that all terms proportional to $\delta^2$ have been neglected.

The stability of the fixed point solutions is therefore determined by the eigenvalues of the resulting perturbation matrix. A lengthy calculation which
we will omit here shows that the point $(0,0,0)$ leads to the eigenvalues
\begin{equation}
\lambda_1 = \frac{3(\omega-1)}{2}, \hspace{0.5cm} \lambda_2 = \frac{3(1+\omega)}{2}, \hspace{0.5cm} \lambda_3 = \frac{3(1+\omega)}{2}
\end{equation}
which indicates that this is never a point of stability for the theory unless the equation of state is phantom ie $\omega < -1$. In fact this statement will
be true for all the various cases we have considered in the physical limit, since the dynamical equations of motion all reduce to the exact same form in
this instance.

Another relatively simple case to consider is that in Case III.
For slices through the ($x, y$) plane at $z = 0$ we find the
eigenvalues
\begin{eqnarray}
\lambda &=& \frac{1}{2}(x^2 + y^2) + \frac{3}{2}\left(\omega(x^2 -1) -1\right)  \\
\lambda_{\pm} & = & \frac{1}{4x^2}\left(-6x^4 (1+\omega) -6x^2 + 2 y^2 x^2 +5y^2 \pm F(x,y) \right) \nonumber \\
F(x,y) &=& \sqrt{12y^2 x^2 \omega + 96 y^2 x^4 \omega - 48 y^2 x^6 \omega - 8 y^4 x^2 + 48 y^2 x^4 + 16 y^4 x^4 - 48y^2x^6 + 36\omega^2 x^4
+17y^4}\,. \nonumber
\end{eqnarray}
If one now slices this through $y = 0$ we see that we are left with the same situation discussed above (as expected), indicative of a phantom
equation of state.

On the other hand, through the $y = 0$ plane we see that the
eigenvalues become
\begin{eqnarray}
\lambda &=& \frac{3}{2}(1 + \omega)\left(1-z^2 -x^2 \left( 1 - \frac{Q z^2}{x^2} \right)  \right)   \\
\lambda_{\pm} & = & -\frac{3}{2x}\left( -Qz^2 -x + 2 x z^2 + x^3 - x z^2 Q + x^2 \pm F(x,y)  \right) \nonumber
\end{eqnarray}
where $F$ is another polynomial in $x, z$ and we have defined $Q = T/\tilde V$ for simplicity. In the limit that
 $z \rightarrow 0$, we find that these simplify to yield
\begin{eqnarray}
\lambda &\rightarrow& \frac{3}{2}(1 + \omega)(1- x^2)\,,   \nonumber \\
\lambda_{\pm} & \rightarrow & \frac{3}{2}(1 + \omega)\left(x^2 -1 -x^2(1\pm 1) \right).
\end{eqnarray}
Note that two of the eigenvalues are therefore degenerate as before, requiring
 a phantom equation of state, however the final eigenvalue has the
opposite sign and therefore this fixed point is always unstable.

 The remaining fixed points can be analyzed in precisely the same manner, although the analysis is somewhat awkward. We will
postpone the relevant discussion here and return to it in a follow-up publication.
\section{Discussion}
We have initiated an alternate approach to the problem of k-essence, or DBI quintessence \cite{Martin:2008xw}, 
using a more generalised form of the DBI action. Since this has more
degrees of freedom, the resulting analysis is typically complicated, but the phase space structure is far richer.
We have attempted to make some headway by restricting the phase space volume to various two-dimensional slices,
and attempting to identify the relevant solution curves upon which the fixed points may lie. Our ansatz for each
of the unknown functions is also potentially restrictive, however we are confident that it represents the leading semi-classical contributions which may (or may-not)
be derivable from a full string theory embedding of out model.

What is clear is that the ratio of the (warped) brane tension to the potential is an important factor in the dynamics of the theory, where we
found $T \ge \tilde V$ in several cases. Moreover the additional multiplicative factor $W(\phi)$ plays a crucial role, even when it is a
constant, since it comes into the field equations non-trivially in the expression for $H'/H$. In the usual DBI analysis, $W=1$ and the
tension is the sole term responsible for the interesting quintessence behaviour. In some string compactifications,
where the warp factor has no cut-off at small distances, we typically find $W$ is constant and greater than
unity. However there may be entire classes of solution where $W \le 1$, which can lead to novel phase space trajectories.
Since our approach has been phenomenological, and that there may be additional string backgrounds of interest that have yet to be fully
explored, we cannot rule out $W < 1$ - which is vital for obtaining fixed point solutions in Case I for example.

Our numerical results have shown that there is indeed a rich phase space structure present due to the increased number of degrees of freedom. We expect many
of these to yield highly non-trivial stable fixed points in the full analysis, which is beyond the scope of the current note. We have classified
the nature of as many of the fixed points as is feasible within the current analysis.
Ultimately we hope that this will lead to a renewed interest in dynamical dark energy models driven by a more generalised approach to $D3$-brane dynamics.

In light of the recent developments in holographic dark energy \cite{Li:2004rb, Zhang:2005hs} and the apparent relation to agegraphic \cite{Cai:2007us, Wei:2007ty}
dark energy, we hope that it may be possible to reconstruct the various potentials in our generalised model along the lines of \cite{Cui:2009ns}.
\section*{Acknowledgments} B.G. is supported by
the Thailand Research Fund and the Commission on Higher Education. In Cambridge, B.G. is sponsored by Naresuan University's Overseas
Postdoctoral Research Fellowship and by the Centre for Theoretical Cosmology, D.A.M.T.P., University of Cambridge at which much gratitude is
expressed to Anne-Christine Davis, Stephen Hawking and Neil Turok for support. A special thank is given to Chakkrit Kaeonikhom for editing the
figures, and to Sudhakar Panda for useful comments.



\begin{thebibliography}{99}

\bibitem{Spergel:2006hy}
  D.~N.~Spergel {\it et al.}  [WMAP Collaboration],
  Astrophys.\ J.\ Suppl.\  {\bf 170}, 377 (2007)
  [arXiv:astro-ph/0603449].

\bibitem{Komatsu:2008hk}
  E.~Komatsu {\it et al.}  [WMAP Collaboration],
  arXiv:0803.0547 [astro-ph].

\bibitem{Riess:1998cb}
  A.~G.~Riess {\it et al.}  [Supernova Search Team Collaboration],
  Astron.\ J.\  {\bf 116}, 1009 (1998)
  [arXiv:astro-ph/9805201].

\bibitem{Tegmark:2003ud}
  M.~Tegmark {\it et al.}  [SDSS Collaboration],
  Phys.\ Rev.\  D {\bf 69}, 103501 (2004)
  [arXiv:astro-ph/0310723].


\bibitem{Grana:2005jc}
  M.~Grana,
  Phys.\ Rept.\  {\bf 423}, 91 (2006)
  [arXiv:hep-th/0509003].


\bibitem{Conlon:2005jm}
  J.~P.~Conlon and F.~Quevedo,
  JHEP {\bf 0601}, 146 (2006)
  [arXiv:hep-th/0509012].
  M.~Cicoli, C.~P.~Burgess and F.~Quevedo,
  arXiv:0808.0691 [hep-th].


\bibitem{Copeland:2006wr}
  E.~J.~Copeland, M.~Sami and S.~Tsujikawa,
  Int.\ J.\ Mod.\ Phys.\  D {\bf 15}, 1753 (2006)
  [arXiv:hep-th/0603057].
  B.~Gumjudpai, T.~Naskar, M.~Sami and S.~Tsujikawa,
  JCAP {\bf 0506}, 007 (2005)
  [arXiv:hep-th/0502191].

\bibitem{Peebles:1987ek}
  P.~J.~E.~Peebles and B.~Ratra,
  Astrophys.\ J.\  {\bf 325}, L17 (1988).
  B.~Ratra and P.~J.~E.~Peebles,
  Phys.\ Rev.\  D {\bf 37}, 3406 (1988).
  A.~R.~Liddle and R.~J.~Scherrer,
  Phys.\ Rev.\  D {\bf 59}, 023509 (1999)
  [arXiv:astro-ph/9809272].
  I.~Zlatev, L.~M.~Wang and P.~J.~Steinhardt,
  Phys.\ Rev.\ Lett.\  {\bf 82}, 896 (1999)
  [arXiv:astro-ph/9807002].

\bibitem{ArmendarizPicon:2000dh}
  C.~Armendariz-Picon, V.~F.~Mukhanov and P.~J.~Steinhardt,
  Phys.\ Rev.\ Lett.\  {\bf 85}, 4438 (2000)
  [arXiv:astro-ph/0004134].
  C.~Armendariz-Picon, V.~F.~Mukhanov and P.~J.~Steinhardt,
  Phys.\ Rev.\  D {\bf 63}, 103510 (2001)
  [arXiv:astro-ph/0006373].
  R.~R.~Caldwell,
  Phys.\ Lett.\  B {\bf 545}, 23 (2002)
  [arXiv:astro-ph/9908168].

\bibitem{Sen:2002in}
  A.~Sen,
  JHEP {\bf 0207}, 065 (2002)
  [arXiv:hep-th/0203265].
  A.~Sen,
  JHEP {\bf 0204}, 048 (2002)
  [arXiv:hep-th/0203211].
  G.~W.~Gibbons,
  Phys.\ Lett.\  B {\bf 537}, 1 (2002)
  [arXiv:hep-th/0204008].
  A.~Sen,
  JHEP {\bf 9910}, 008 (1999)
  [arXiv:hep-th/9909062].
  J.~Kluson,
  Phys.\ Rev.\  D {\bf 62}, 126003 (2000)
  [arXiv:hep-th/0004106].
  E.~A.~Bergshoeff, M.~de Roo, T.~C.~de Wit, E.~Eyras and S.~Panda,
  JHEP {\bf 0005}, 009 (2000)
  [arXiv:hep-th/0003221].
  M.~R.~Garousi,
  Nucl.\ Phys.\  B {\bf 584}, 284 (2000)
  [arXiv:hep-th/0003122].

\bibitem{Alishahiha:2004eh}
  M.~Alishahiha, E.~Silverstein and D.~Tong,
  Phys.\ Rev.\  D {\bf 70}, 123505 (2004)
  [arXiv:hep-th/0404084].
  E.~Silverstein and D.~Tong,
  Phys.\ Rev.\  D {\bf 70}, 103505 (2004)
  [arXiv:hep-th/0310221].

\bibitem{Maldacena:2002vr}
  J.~M.~Maldacena,
  JHEP {\bf 0305}, 013 (2003)
  [arXiv:astro-ph/0210603].
  X.~Chen, M.~x.~Huang, S.~Kachru and G.~Shiu,
  JCAP {\bf 0701}, 002 (2007)
  [arXiv:hep-th/0605045].


\bibitem{Martin:2008xw}
  J.~Martin and M.~Yamaguchi,
  Phys.\ Rev.\  D {\bf 77}, 123508 (2008)
  [arXiv:0801.3375 [hep-th]].

\bibitem{Guo:2008sz}
  Z.~K.~Guo and N.~Ohta,
  JCAP {\bf 0804}, 035 (2008)
  [arXiv:0803.1013 [hep-th]].

\bibitem{Kecskemeti:2006cg}
  S.~Kecskemeti, J.~Maiden, G.~Shiu and B.~Underwood,
  JHEP {\bf 0609}, 076 (2006)
  [arXiv:hep-th/0605189].
  X.~Chen,
  Phys.\ Rev.\  D {\bf 71}, 063506 (2005)
  [arXiv:hep-th/0408084].
  J.~Ward,
  JHEP {\bf 0712}, 045 (2007)
  [arXiv:0711.0760 [hep-th]].
  X.~Chen,
  arXiv:0807.3191 [hep-th].
  F.~Gmeiner and C.~D.~White,
  JCAP {\bf 0802}, 012 (2008)
  [arXiv:0710.2009 [hep-th]].

\bibitem{Easson:2007dh}
  D.~A.~Easson, R.~Gregory, D.~F.~Mota, G.~Tasinato and I.~Zavala,
  JCAP {\bf 0802}, 010 (2008)
  [arXiv:0709.2666 [hep-th]].
\bibitem{Huang:2007hh}
  M.~x.~Huang, G.~Shiu and B.~Underwood,
  Phys.\ Rev.\  D {\bf 77}, 023511 (2008)
  [arXiv:0709.3299 [hep-th]].

\bibitem{Cai:2008if}
  Y.~F.~Cai and W.~Xue,
  arXiv:0809.4134 [hep-th].

\bibitem{Cai:2009hw}
  Y.~F.~Cai and H.~Y.~Xia,
  arXiv:0904.0062 [hep-th].


\bibitem{Langlois:2009ej}
  D.~Langlois, S.~Renaux-Petel and D.~A.~Steer,
  arXiv:0902.2941 [hep-th].

\bibitem{Avgoustidis:2006zp}
  A.~Avgoustidis, D.~Cremades and F.~Quevedo,
  Gen.\ Rel.\ Grav.\  {\bf 39}, 1203 (2007)
  [arXiv:hep-th/0606031].
  A.~Avgoustidis and I.~Zavala,
  arXiv:0810.5001 [hep-th].

\bibitem{Lidsey:2007gq}
  J.~E.~Lidsey and I.~Huston,
  JCAP {\bf 0707}, 002 (2007)
  [arXiv:0705.0240 [hep-th]].
  D.~Baumann and L.~McAllister,
  Phys.\ Rev.\  D {\bf 75}, 123508 (2007)
  [arXiv:hep-th/0610285].

\bibitem{Thomas:2007sj}
  S.~Thomas and J.~Ward,
  Phys.\ Rev.\  D {\bf 76}, 023509 (2007)
  [arXiv:hep-th/0702229].
  S.~Thomas and J.~Ward,
  JHEP {\bf 0610}, 039 (2006)
  [arXiv:hep-th/0508085].

\bibitem{Ashoorioon:2009wa}
  A.~Ashoorioon, H.~Firouzjahi and M.~M.~Sheikh-Jabbari,
  arXiv:0903.1481 [hep-th].


\bibitem{Alabidi:2008ej}
  L.~Alabidi and J.~E.~Lidsey,
  arXiv:0807.2181 [astro-ph].

\bibitem{Huston:2008ku}
  I.~Huston, J.~E.~Lidsey, S.~Thomas and J.~Ward,
  JCAP {\bf 0805}, 016 (2008)
  [arXiv:0802.0398 [hep-th]].

\bibitem{Becker:2007ui}
  M.~Becker, L.~Leblond and S.~E.~Shandera,
  Phys.\ Rev.\  D {\bf 76}, 123516 (2007)
  [arXiv:0709.1170 [hep-th]].
  T.~Kobayashi, S.~Mukohyama and S.~Kinoshita,
  JCAP {\bf 0801}, 028 (2008)
  [arXiv:0708.4285 [hep-th]].
  S.~Mukohyama,
  arXiv:0706.3214 [hep-th].

\bibitem{Silverstein:2008sg}
  E.~Silverstein and A.~Westphal,
  Phys.\ Rev.\  D {\bf 78}, 106003 (2008)
  [arXiv:0803.3085 [hep-th]].
L.~McAllister, E.~Silverstein and A.~Westphal,
arXiv:0808.0706 [hep-th].

\bibitem{Leblond:2008gg}
  L.~Leblond and S.~Shandera,
  JCAP {\bf 0808}, 007 (2008)
  [arXiv:0802.2290 [hep-th]].

\bibitem{Myers:1999ps}
  R.~C.~Myers,
  JHEP {\bf 9912}, 022 (1999)
  [arXiv:hep-th/9910053].
  A.~A.~Tseytlin,
  Nucl.\ Phys.\  B {\bf 501}, 41 (1997)
  [arXiv:hep-th/9701125].

\bibitem{Li:2004rb}
  M.~Li,
  Phys.\ Lett.\  B {\bf 603}, 1 (2004)
  [arXiv:hep-th/0403127].

\bibitem{Zhang:2005hs}
  X.~Zhang and F.~Q.~Wu,
  Phys.\ Rev.\  D {\bf 72}, 043524 (2005)
  [arXiv:astro-ph/0506310].
  X.~Zhang and F.~Q.~Wu,
  Phys.\ Rev.\  D {\bf 76}, 023502 (2007)
  [arXiv:astro-ph/0701405].
  Q.~G.~Huang and Y.~G.~Gong,
  JCAP {\bf 0408}, 006 (2004)
  [arXiv:astro-ph/0403590].
  Q.~Wu, Y.~Gong, A.~Wang and J.~S.~Alcaniz,
  Phys.\ Lett.\  B {\bf 659}, 34 (2008)
  [arXiv:0705.1006 [astro-ph]].

\bibitem{Cai:2007us}
  R.~G.~Cai,
  Phys.\ Lett.\  B {\bf 657}, 228 (2007)
  [arXiv:0707.4049 [hep-th]].

\bibitem{Wei:2007ty}
  H.~Wei and R.~G.~Cai,
  Phys.\ Lett.\  B {\bf 660}, 113 (2008)
  [arXiv:0708.0884 [astro-ph]].

\bibitem{Cui:2009ns}
  J.~Cui, L.~Zhang, J.~Zhang and X.~Zhang,
  arXiv:0902.0716 [astro-ph.CO].




\end{thebibliography}
\end{document}